\documentclass[amsmath,amssymb,superscriptaddress,floatfix]{revtex4-1}
\usepackage{graphicx}
\usepackage{dcolumn}
\usepackage{bm}
\usepackage{amsfonts}
\usepackage{amsmath}
\usepackage{ulem}
\usepackage{color}
\usepackage{hyperref}
\begin{document}
\title{Photoexcitation in two-dimensional topological insulators: \\
Generating and controlling electron wavepackets in Quantum Spin Hall systems}
\author{Fabrizio Dolcini}
\email{fabrizio.dolcini@polito.it}
\affiliation{Dipartimento di Scienza Applicata e Tecnologia, Politecnico di Torino\\ corso Duca degli Abruzzi 24, 10129 Torino (Italy)}

\author{Fausto Rossi}
\affiliation{Dipartimento di Scienza Applicata e Tecnologia, Politecnico di Torino\\ corso Duca degli Abruzzi 24, 10129 Torino (Italy)}

\begin{abstract}
One of the most fascinating challenges in Physics is the realization of an electron-based counterpart of quantum optics, which requires the capability to generate and control single electron wave packets. The edge states of quantum spin Hall (QSH) systems, i.e. two-dimensional (2D) topological insulators realized in HgTe/CdTe and InAs/GaSb quantum wells, may turn the tide in the field, as they do not require the magnetic field that limits the implementations based on quantum Hall effect. However, the  band structure of these topological  states, described by a massless Dirac fermion Hamiltonian, prevents electron photoexcitation via the customary vertical electric dipole transitions of conventional optoelectronics. So far, proposals to overcome this problem are based on magnetic dipole transitions induced via Zeeman coupling by circularly polarised radiation, and are limited by the $g$-factor. Alternatively, optical transitions can be induced from the edge states to the bulk states, which are not topologically protected though.  

\indent Here  we show   that   an electric pulse, localized in space and/or time and applied at  a QSH edge, can photoexcite electron wavepackets by intra-branch electrical transitions, without invoking the bulk states or the Zeeman coupling. Such wavepackets are  spin-polarised and  propagate in opposite directions, with a density profile that is independent of the    initial equilibrium temperature and that does not exhibit dispersion, as a result of  the linearity of the spectrum and of the chiral anomaly characterising massless Dirac electrons. We also investigate the photoexcited energy distribution and show  how, under appropriate circumstances, minimal excitations (Levitons) are generated. Furthermore, we show that the presence of a Rashba spin-orbit coupling can be exploited to tailor the shape of photoexcited wavepackets. Possible experimental realizations are also discussed.
\end{abstract}
\maketitle
\section{Introduction}
\label{sec:1}
Electron quantum optics is one of the most  fascinating and rapidly growing fields in Physics. Its goal, i.e. the realization of an electron-based counterpart of quantum optics,   requires the capability to generate and  control single electron wave packets, where one could encode, transmit and process  information~\cite{bertoni2000,ritchie_2007,bocquillon2012}.
Presently,  the edge channels  of quantum Hall (QH) systems  are  considered the reference platform to achieve this purpose \cite{feve2007,roulleau2008,glattli2010,feve2011,bocquillon2013,bocquillon_science_2013,kataoka2013,bocquillon2014,martin_2014b,feve2015,kataoka2015,janssen2016}: electrons  injected by electron  pumps  in these chiral one-dimensional (1D) edge channels  can propagate ballistically and coherently over various micrometers, topologically protected from  backscattering off disorder.
Furthermore, quantum point contacts can be used as the analogue of electron beam splitters.  
There is, however, a major drawback limiting the large scale applications of QH-based electron quantum optics, namely the strong values of magnetic field   that are needed to generate the ballistic edge states.

A quite promising alternative is the quantum spin Hall (QSH) effect\cite{kane-mele2005a,kane-mele2005b,bernevig_science_2006}, where edge channels do not require any applied magnetic field, as they originate  from a spin-orbit induced topological transition. These QSH edge channels,
observed in   HgTe/CdTe\cite{konig_2006,molenkamp-zhang_jpsj,roth_2009,brune_2012} and InAs/GaSb\cite{liu-zhang_2008,knez_2007,knez_2014,spanton_2014} quantum wells, are also protected from backscattering off non-magnetic impurities, as their  group velocity is locked to their spin orientation. 
Notably, QSH edge states also offer two more important advantages. First,  quite similarly to photons, they exhibit a linear electronic spectrum, where the Fermi velocity $v_F$ plays the role of the speed of light $c$. As a consequence, a freely propagating electronic wave packet does not feature the customary dispersion arising in conventional parabolic band materials.
This is crucial for the information transmission rate, which requires  the generation of  wavepackets sequences that propagate  without overlapping. 
Second, QSH edge states are helical, meaning that their spin orientation is locked to the group velocity. This means that electron wavepackets propagating in a given direction along the edge are characterized by a well defined spin polarization.    
For all these reasons, QSH edge states may be a promising platform for electron quantum optics\cite{bercioux_2013,buttiker_2013,martin_2014,recher_2015,sassetti_2016,dolcini_2016,dolcini_2017}.

In order to generate electron wave packets in a controlled way, photoexcitation is perhaps the most customary approach. However, QSH edge states exhibit an important difference with respect to customary optoelectronics systems: the vertical electric dipole transitions that typically occur between valence and conduction bands of conventional semiconductors are forbidden in QSH edge states, due to a selection rule arising from their helical nature.
To tackle this problem, some works have proposed to exploit circularly polarised radiations, whose magnetic field can induce magnetic dipole transitions on the edge states. This approach, relying on   Zeeman coupling\cite{cayssol2012,artemenko2013,dolcetto-sassetti2014} is, however, limited by the rather small  $g$-factor.  
Alternatively,   it is possible to induce optical  transitions from the edge states to the bulk states~\cite{artemenko2013,artemenko2015} thereby losing, however, the topological  protection from disorder, which is one of the most interesting features of topological systems.  
Most of these approaches are based on what is known in optoelectronics as the {\it far field} regime, where a monochromatic radiation is applied over the whole sample for a duration that is long compared to its oscillation period.\\

In this article, we show that in the so called {\it near field} regime the photoexcitation in QSH edge states is possible without invoking the bulk states or the Zeeman coupling:  when the electric field is applied on a spatially {\it localised} region and for a {\it finite time}, one can photoexcite localised electron wave packets that propagate with a density profile that maintains its shape unaltered, without dispersion and with a well specified spin polarization.  Such photoexcited space density profile is independent of the temperature of the initial equilibrium state and depends only -- and linearly-- on the intensity of the applied pulse. These properties, which are derived exactly, are due to the linearity of the spectrum and the chiral anomaly effect characterizing QSH edge states, and do not rely on any  linear response assumption.
We also show that  the energy correlations induced by the photoexcitation  depend on the initial temperature,  and that the  photoexcited energy distribution   does not merely amount to particle-hole excitations, i.e. to a reshuffling  of electron states from below to above the equilibrium Fermi level as a result of the electric pulse. Instead, due to the chiral anomaly effect, the electric pulse effectively gives rise to a net creation of charge of one electron branch (compensated by the annihilation of charges of the opposite branch) and  under appropriate circumstances  one can generate minimal (i.e. purely particle or purely hole) excitations in each QSH helical branch. So far, these excitations, also known as Levitons\cite{levitov1996,levitov1997,levitov2006,moskalets2014,flindt2015,moskalets2016,moskalets2017,martin-sassetti_PRL_2017,martin-sassetti_PRB_2017}, have been  observed in ballistic channels in   2DEGs\cite{glattli2013,glattli_PRB_2013,glattli2014,glattli2017}. Furthermore, by analyzing the interplay between the photoexcitation process and the presence of Rashba spin-orbit interaction, we show that the latter can be exploited to tailor the shape of the photoexcited QSH wavepackets.

The article is organised as follows. In Sec.\ref{sec:2} we present the model, briefly recall the goal of the photoexcitation problem, and present the exact solution of the massless Dirac equation coupled to an electromagnetic pulse. Then, in Sec.\ref{sec:3}, we use such result to determine the general expressions for the  photoexcited electron density profile and energy correlations induced by the photoexcitation process,   pointing out how to obtain  gauge-invariant results. In Sec.\ref{sec:4} we specify such general results to various types of applied electric pulse, computing the photoexcited  electron density profile and the photoexcited energy distribution.
  Finally, in Sec.\ref{sec:5} we discuss the interplay between the photoexcitation and the presence of Rashba spin-orbit interaction, whereas in Sec.\ref{sec:6} we propose some possible experimental realization schemes. We finally summarize and conclude in Sec.\ref{sec:7}.

\section{Model for photoexcitation in QSH edge states}
\label{sec:2}
\subsection{Hamiltonian}
Let us focus on one edge of a QSH system and denote by $x$ the coordinate along the boundary. A Kramers' pair of one-dimensional (1D) counterflowing helical states is described by a spinor field operator  $\Psi(x)=(\psi^{}_{\uparrow}(x) \,,\, \psi^{}_{\downarrow}(x) )^T$,
where $\uparrow$ and $\downarrow$ refer to the QSH edge  spin orientation. 
The electronic system, initially in an equilibrium state, is then exposed to an externally applied electric field, expressed as 
\begin{equation}
\label{E-field}
E(x,t)=-\partial_x V- \partial_t A/c\
\end{equation}
in terms of  scalar and vector potential, $V$ and $A$, respectively. Here $c$ is the speed of light.
The full Hamiltonian thus reads
\begin{equation}\label{Hfull}
\hat{\mathcal{H}}  = \hat{\mathcal{H}}_\circ\,\,+\, \hat{\mathcal{H}}_{em}\quad,
\end{equation} 
where $\hat{\mathcal{H}}_\circ$ denotes the electronic contribution, while 
\begin{eqnarray}
\hat{\mathcal{H}}_{em} &= &{\rm e} \int   dx V(x,t)  \, \hat{n}\, -\frac{{\rm e}}{c} \int   dx \, A(x,t) \,\hat{J}\label{Hem} 
\end{eqnarray}
describes the coupling to the electromagnetic field, where
\begin{equation}\label{n-def}
\hat{n}=\Psi^\dagger(x) \Psi(x) = \hat{n}_\uparrow+\hat{n}_\downarrow
\end{equation}
is the total electron density operator, with $\hat{n}_{\sigma}\doteq \psi^\dagger_\sigma\psi^{}_\sigma$ ($\sigma=\uparrow,\downarrow$) denoting the  spin-resolved density, and $\hat{J}$ is the current density operator, whose explicit expression depends on the electronic term $ \hat{\mathcal{H}}_\circ$.  In particular, we shall mostly focus on the case where the electronic contribution describes the purely linear spectrum of the QSH edge states, and consists of the massless Dirac fermion `kinetic' term 
\begin{equation}\label{Hkin}
\hat{\mathcal{H}}_\circ=\hat{\mathcal{H}}_{kin} =v_F \int   dx \, \Psi^\dagger(x) \,   \sigma_3  {p}_x \, \Psi(x)  
\end{equation}
where $v_F$ is the Fermi velocity, $p_x=-i\hbar \partial_x$, and $\sigma_{1,2,3}$  are  Pauli matrices. Then, the expression for the electronic current $\hat{J}$ appearing in (\ref{Hem})  reads
\begin{eqnarray}
\hat{J}&=&v_F \Psi^\dagger(x) \, \sigma_3 \Psi(x)=  v_F\left( \hat{n}_\uparrow-\hat{n}_\downarrow  \right) \label{J-def}
\end{eqnarray}
We anticipate, however, that in Sec.\ref{sec:5}  we shall include in $\hat{\mathcal{H}}_\circ$ also the Rashba spin-orbit coupling along the edge, and in that case the expression of $\hat{J}$ gets modified.
Finally, the   Zeeman coupling ascribed to the magnetic field arising from time-dependence of $E(x,t)$ can be shown to be negligible\cite{dolcini_2017} and will be omitted henceforth.  
\subsection{Statement of the problem and the gauge invariance issue}
In the initial equilibrium state, i.e. before the term (\ref{Hem}) is switched on, the space-time evolution of the electron field operator is dictated by the term (\ref{Hkin}) and is given by
\begin{equation}
\psi^\circ_{\sigma}(x,t)=
\frac{1}{\sqrt{2\pi \hbar v_F}} \int dE e^{-i \frac{E}{\hbar} (t \mp \frac{x}{v_F})} c^\circ_{E,\sigma}   \hspace{1cm} \sigma=\uparrow,\downarrow=\pm \quad,
\end{equation}
where the correlation between the energy mode operators $c^\circ_{E,\sigma}$ 
\begin{eqnarray}\label{Gcirc}
\mathcal{G}^\circ_\sigma(E,\tilde{E})& \doteq& \left\langle c^{\circ\, \dagger}_{E+\frac{\tilde{E}}{2},\sigma} c^\circ_{E-\frac{\tilde{E}}{2},\sigma} \right\rangle =   \delta(\tilde{E}) \, f^\circ(E)\,   
\end{eqnarray}
is purely diagonal and characterized by the Fermi equilibrium distribution function $f^\circ(E)=\left[1+\exp[(E-\mu)/k_B T]\right]^{-1}$, with chemical potential $\mu$ and temperature $T$. Here energies are measured with respect to the Dirac level.  
Furthermore, the space profile of the equilibrium density is uniform 
\begin{equation}
n^\circ_\sigma=\langle \psi^{\circ \, \dagger}_{\sigma}(x,t) \psi^{\circ}_{\sigma}(x,t)\rangle= \mbox{const}= \frac{\mu+E_c}{2\pi \hbar v_F}
\end{equation}
where $E_c$ denotes the ultraviolet energy cut-off of the band dispersion.\\

When the electric pulse  is applied, the system is driven out of equilibrium by the term Eq.(\ref{Hem}), and the space profile as well as the energy correlations are modified. The goal of the investigation is to determine the photoexcited density profile and energy correlations, i.e. the {\it deviations}   of these quantities with respect to their equilibrium value,
\begin{eqnarray}
\Delta n_\sigma(x,t) &\doteq &   \langle \psi^{\dagger}_{\sigma}(x,t) \psi^{}_{\sigma}(x,t)\rangle \,\,-   \langle \psi^{\circ \, \dagger}_{\sigma}(x,t) \psi^{\circ}_{\sigma}(x,t)\rangle  \\
\Delta \mathcal{G}_\sigma(E,\tilde{E};x) &\doteq & \langle c^{\dagger}_{E+\frac{\tilde{E}}{2},\sigma}(x) c^{}_{E-\frac{\tilde{E}}{2},\sigma}(x) \rangle\,-\, \langle c^{\circ\, \dagger}_{E+\frac{\tilde{E}}{2},\sigma} c^\circ_{E-\frac{\tilde{E}}{2},\sigma} \rangle
\end{eqnarray}
where $\psi_{\sigma}(x,t)$ is the electron field evolution for the full Hamiltonian (\ref{Hfull}), and  the local (i.e. $x$-dependent) density mode operators, defined as
\begin{equation}
c_{E,\sigma}(x) \doteq \sqrt{\frac{v_F}{2\pi \hbar}} \int dt \, e^{+i \frac{E}{\hbar} (t \mp \frac{x}{v_F})} \psi_{\sigma}(x,t) \quad,
\end{equation}
identify the energy weight of the electron field operator at the space point $x$.\\

Importantly, we observe that the Hamiltonian (\ref{Hfull}) depends on the specific gauge $(V,A)$ chosen to describe the electric field (\ref{E-field}). However, 
a physically meaningful observable can only depend on the applied electric field, and not on the gauge choice. Thus, the essential  prerequisite for a correct result about photoexcitation   is that it must be left {\it invariant} by any gauge transformation
\begin{equation}\label{gauge-inv}
\left\{ 
\begin{array}{ccl} 
 \displaystyle V & \rightarrow & \displaystyle V^\prime =V-(\hbar/\rm e) \partial_t \chi   \\  
 \displaystyle A & \rightarrow & \displaystyle A^\prime =A+(\hbar c/{\rm e}) \partial_x \chi  \\
 \Psi(x,t) & \rightarrow  & \displaystyle  \Psi^\prime(x,t) = e^{i \chi(x,t)} \Psi(x,t)  
  \end{array}
\right.
\end{equation}
where $\chi$ is any arbitrary function.
We emphasize that, since gauge invariance issue is tightly connected to the conservation of electrical charge, this aspect cannot be overlooked. In systems described by a linear spectrum
 (massless Dirac fermions), such requirement involves some subtlelties that are not present in conventional systems described a parabolic spectrum (Schr\"odinger fermions), as we shall discuss below.

\subsection{Exact solution of the massless Dirac equation coupled to an electric field  in 1+1 dimensions}
\label{sec:1-2}
In a given  gauge $(V,A)$ of electromagnetic potentials, the equations of motion for the electron field operator are dictated by the Hamiltonian (\ref{Hfull}) with Eqs.(\ref{Hem}) and (\ref{Hkin}), and read
\begin{equation}\label{eom-Psi}
i\hbar \partial_t  \Psi  =\left( v_F \sigma_3\, (\hat{p} -\frac{\rm e}{c} A(x,t)) +{\rm e}V(x,t) \sigma_0 \right)\, \Psi \quad,
\end{equation}
where $\sigma_0$ is the $2\times 2$ identity matrix. Equation (\ref{eom-Psi}) is the massless Dirac equation coupled to the electromagnetic potentials in 1+1 dimensions. Since the matrix term on the right-hand side    is purely diagonal, the equations for the two components $\psi_\uparrow$ and $\psi_\downarrow$ are decoupled and  can be solved exactly \cite{dolcini_2016}, obtaining
\begin{equation}\label{Psi-sol-gen}
\psi_{\uparrow,\downarrow}(x,t)=  \psi^\circ_{\uparrow,\downarrow}(x \mp v_F t) \,  e^{\pm i \phi_{\uparrow,\downarrow}(x,t)} \,  \quad,
\end{equation}
where $\psi^\circ_{\uparrow,\downarrow}(x \mp v_F t)$ denotes  the space-time evolution of the electron field component  of Eq.(\ref{Hkin}), i.e. in absence of the electromagnetic coupling, and represents genuine right-moving electrons with spin-$\uparrow$ and left-moving electrons with spin-$\downarrow$, respectively. In contrast, the phases  $\phi_{\uparrow,\downarrow}$   encode  the effect of the electromagnetic  potentials $V(x,t)$ and $A(x,t)$, and can be given two equivalent expressions
\begin{eqnarray} 
\phi_{\uparrow}(x,t)    & =  &   \frac{{\rm e} v_F}{\hbar c} \int_{-\infty}^t \!\! (A-\frac{c}{v_F}V)(x-v_F (t-t^\prime), t^\prime) \, dt^\prime = \nonumber \, \,  \\ 
&= & \frac{{\rm e}}{\hbar c} \int_{-\infty}^x \, (A-\frac{c}{v_F}V)(x^\prime, t-\frac{x-x^\prime}{v_F}) \, dx^\prime  \label{phi-up} 
\end{eqnarray}
and
\begin{eqnarray}
\phi_{\downarrow}(x,t)    &= &  \displaystyle  \frac{{\rm e} v_F}{\hbar c} \int_{-\infty}^t \!\! (A+\frac{c}{v_F}V)(x+v_F (t-t^\prime), t^\prime) \, dt^\prime = \nonumber  \\ 
&= & \displaystyle   \frac{\rm e}{\hbar c} \int_{x}^\infty \, (A+\frac{c}{v_F}V)(x^\prime, t+\frac{x-x^\prime}{v_F}) \, dx^\prime \quad. \label{phi-dn}
\end{eqnarray}
The above expressions have a straightforward physical interpretation. The first (second) line of Eq.(\ref{phi-up}), for instance,   expresses the phase $\phi_\uparrow(x,t)$ induced by the electromagnetic field on the helical right-moving electron $\psi^\circ_{\uparrow}(x-v_F t)$ as a convolution over time (space) of the values of the electromagnetic potentials at times earlier than~$t$ and at positions located on the left of~$x$,   propagating with the electron Fermi velocity $v_F$ according to the   dynamics dictated by Eq.(\ref{Hkin}). 
It is worth pointing out that the phases (\ref{phi-up})-(\ref{phi-dn}) are gauge {\it dependent}, as they characterize the exact solution of the gauge-dependent equation of motion (\ref{eom-Psi}). In the next section we shall show how to obtain gauge {\it independent} results.

\section{General results for photoexcitation}
\label{sec:3}
\subsection{Photoexited electron density profiles}
\label{sec:3:1}
Since the whole effect of the electromagnetic coupling amounts to the {\it phases} (\ref{phi-up})-(\ref{phi-dn})  that multiply  the result $\psi_\sigma^\circ$ in absence of electric pulse, one might at first naively think that such phases  drop out when computing field bilinear from Eq.(\ref{Psi-sol-gen}), such as densities $\hat{n}_\sigma=\psi^\dagger_\sigma \psi^{}_\sigma$ (with $\sigma=\uparrow,\downarrow$). That would mean that the numbers of right- and left-moving electrons are separately conserved,  remaining equal to the value $\hat{n}^\circ_\sigma= {\psi^\circ_\sigma}^\dagger {\psi^\circ_\sigma}^{}$  without electric field. One would thus be tempted to  conclude that the electric pulse does not alter   the  equilibrium density profiles ($\Delta \hat{n}_\sigma \equiv 0$) and, in particular,  that the {\it chiral density} $\hat{n}^a=\Psi^\dagger \sigma_3\Psi=\hat{n}_\uparrow-\hat{n}_\downarrow$  is conserved. This conservation of the chiral density would also seem to agree with N\"other's theorem,  following from the existence of the chiral symmetry 
\begin{equation}\label{chiral-inv}
\left\{ 
\begin{array}{l} \Psi(x,t)  \rightarrow   \displaystyle  \Psi^\prime(x,t) = e^{i \zeta(x,t) \sigma_3} \Psi(x,t)     
\\     
\begin{array}{lcl}
 \displaystyle V & \rightarrow & \displaystyle V^\prime =V-(\hbar/{\rm e} )\, \sigma_3 \, \partial_t \zeta\\  \displaystyle A & \rightarrow & \displaystyle A^\prime =A+(\hbar c/{\rm e})\,  \sigma_3  \,\partial_x  \zeta \,
\end{array}
\end{array}
\right.
\end{equation}
of the equation of motion (\ref{eom-Psi}). However, such conclusion  is wrong. The obvious physical reason is that an applied electric field must modify the electron current (\ref{J-def}). The mathematical reason is that, despite the existence of such symmetry and the decoupling of the equations (\ref{eom-Psi}), the infinite number of states characterizing the Fermi sea leads to an anomalous breaking of N\"other's conservation law. This non-trivial effect is known as the chiral anomaly. Here we shall only summarize   the main technical aspects related to the 1+1 dimensional case of QSH edge states (details can be found in  Ref.\cite{dolcini_2016}), before describing its effects on the photoexcitation properties.
A  physically correct photoexcited density must be {\it finite} and {\it gauge independent}. 
In order to fulfill these two requirements, one computes the photoexcited density  as
\begin{eqnarray}
 \Delta n_{\sigma}(x,t)  \doteq  
 \!\! \lim_{(x^\prime,t^\prime) \rightarrow (x,t) } \left\langle  \psi^\dagger_{\sigma}(x^\prime,t^\prime)\psi^{}_{\sigma}(x,t)\, e^{-iW_L(x,t,x^\prime,t^\prime)}  \, \, - {\psi^\circ_{\sigma}}^\dagger(x^\prime,t^\prime){\psi^\circ_{\sigma}}^{}(x,t)  \right\rangle_\circ , \hspace{0.2cm} \label{Deltan_{pm}-def-gen} 
\end{eqnarray}
where $\psi_{\sigma}(x,t)$ is the electron field operator   in the presence of the electromagnetic field, Eq.(\ref{Psi-sol-gen}), while $\psi^\circ_{\sigma}(x,t)$ is the one  in the absence of the electromagnetic field. The point-splitting procedure, which is equivalent to introducing a cut-off in $k$-space, enables one to regularize the infinite ground state contribution, which is already present without electromagnetic field and can thus be safely subtracted, thereby obtaining a finite result. However, such result would be gauge-dependent since the point-splitting procedure breaks the gauge invariance. Thus, 
 the Wilson line 
\begin{equation}
W_L(x,t,x^\prime,t^\prime)=\frac{\rm e}{\hbar c} \int_{(x,t)}^{(x^\prime,t^\prime)} (c V d  t^{\prime\prime}-A \, dx^{\prime\prime}) \label{WL-def}
\end{equation}
connecting the two split points $(x^\prime,t^\prime)$ and $(x,t)$ is introduced in Eq.(\ref{Deltan_{pm}-def-gen})  to restore  the gauge invariance, guaranteeing that the result is independent of the specific gauge $(V,A)$ chosen for the electric field (\ref{E-field}).
Applying the above procedure, and inserting the exact solution (\ref{Psi-sol-gen}) into Eq.(\ref{Deltan_{pm}-def-gen}), one obtains the following equivalent expressions for the photoexcited density profiles \cite{dolcini_2016}
\begin{eqnarray}
\Delta n_{\uparrow}(x,t)      
&=& \displaystyle +\frac{\rm e}{2\pi\hbar}   \int_{-\infty}^{t}   \, E(x-v_F (t-t^\prime), t^\prime)  \,   dt^\prime = \nonumber\\  
&=& \displaystyle+\frac{{\rm e}}{2\pi\hbar  v_F}   \int_{-\infty}^{x}   \, E(x^\prime, t-\frac{x-x^\prime}{v_F})  \,   dx^\prime=  \nonumber \\
&=& \displaystyle  + \frac{1}{2\pi}  \left( \partial_x \phi_{\uparrow}(x,t) - \frac{\rm e}{\hbar c} A(x,t) \right)   =  \nonumber \\
&=& \displaystyle  - \frac{1}{2\pi v_F}  \left(  \partial_t \phi_{\uparrow}(x,t)+  \frac{\rm e}{\hbar } V(x,t) \right)   \label{Deltan_{up}_res-gen}
\end{eqnarray}
and
\begin{eqnarray}
\Delta n_{\downarrow}(x,t)  &=&  \displaystyle-\frac{\rm e}{2\pi\hbar}   \int_{-\infty}^{t}   \, E(x+v_F (t-t^\prime), t^\prime)  \,   dt^\prime \, = \nonumber\\ 
&=& \displaystyle -\frac{{\rm e}}{2\pi\hbar  v_F}   \int_{x}^{\infty}   \, E(x^\prime, t+\frac{x-x^\prime}{v_F})  \,   dx^\prime = \nonumber\\
&=& \displaystyle  + \frac{1}{2\pi}  \left( \partial_x \phi_{\downarrow}(x,t) + \frac{\rm e}{\hbar c} A(x,t) \right)   =\nonumber  \\
&=& \displaystyle  + \frac{1}{2\pi v_F}  \left(  \partial_t \phi_{\downarrow}(x,t)-  \frac{\rm e}{\hbar } V(x,t) \right) \quad.
\label{Deltan_{dn}_res-gen}
\end{eqnarray}
Three comments are in order about this result. First, their gauge invariance  clearly appears from the first two lines of Eqs.(\ref{Deltan_{up}_res-gen}) and (\ref{Deltan_{dn}_res-gen}), which depend only on the electric field. Second, they are exact (as far as the model for linear spectrum holds) and do not rely on any linear response approximation. Third, the profile are independent of the temperature $T$ and of the chemical potential $\mu$ of the initial equilibrium state.
\subsubsection{Chiral anomaly}
By taking time and space derivatives of the first lines of the obtained photoexcited densities Eqs.(\ref{Deltan_{up}_res-gen})-(\ref{Deltan_{dn}_res-gen}), one can straightforwardly prove that  
\begin{equation}
\begin{array}{l}
(\partial_t + v_F \partial_x)  \!\Delta   \hat{n}_{\uparrow}(x,t)   \, \,  =  + \frac{\rm e}{2\pi\hbar}  E(x,t)  \, \,     \\
(\partial_t - v_F \partial_x)  \!\Delta   \hat{n}_{\downarrow}(x,t)   \, \,  =  - \frac{\rm e}{2\pi\hbar}  E(x,t)   \, \, ,
\end{array}\label{cont-right-left-new}
\end{equation}
showing that  the electric field $E(x,t)$ breaks the conservation laws appearing on the left hand side, thereby  effectively creating and destroying electrons  in each branch.  Importantly, by taking  sum and difference of Eqs.(\ref{cont-right-left-new}) one finds
\begin{eqnarray} 
\partial_t \Delta\hat{n}(x,t)+\partial_x \Delta\hat{j}(x,t)=0 \hspace{2cm} \label{charge-cons}  \\
\partial_t \Delta\hat{n}^a(x,t)+ \partial_x \Delta\hat{j}^a(x,t)=\frac{\rm e}{\pi\hbar}  E(x,t) \quad.
\label{eq-axial-anomaly}
\end{eqnarray}
The continuity equation (\ref{charge-cons}) for the electron charge $\Delta\hat{n}=\Delta   \hat{n}_{\uparrow}+\Delta   \hat{n}_{\downarrow}$ is fulfilled, due to the gauge invariance under Eq.(\ref{gauge-inv}). However, the conservation law involving the axial charge $\Delta\hat{n}^a=\Delta   \hat{n}_{\uparrow}-\Delta   \hat{n}_{\downarrow}$ and axial current $\Delta\hat{j}^a=v_F (\Delta   \hat{n}_{\uparrow}+\Delta   \hat{n}_{\uparrow})$, which would be expected from the chiral symmetry Eq.(\ref{chiral-inv}), is broken by the anomalous term appearing in Eq.(\ref{eq-axial-anomaly}). This is the chiral anomaly effect,   first discovered in high energy physics~\cite{adler1969,bell-jackiw1969,nielsen1983,bertlmann} and
nowadays on the spotlight in condensed matter physics\cite{burkov2012,li2013,wishvanath2014,takane2016} both in 3D Weyl semimetals~\cite{chen2014,hasan2015a,hasan2015b,ong2015} and in 1D QSH edge states\cite{trauzettel2016}. 
Notably, the anomalous term on the r.h.s. of Eq.(\ref{eq-axial-anomaly}) [or equivalently in Eqs.(\ref{cont-right-left-new})] depends {\it only} on the universal constant ${\rm e}/2\pi \hbar$ and the electric field $E(x,t)$, and not on the electron degrees of freedom. This shows the close relation between the chiral anomaly and the above  mentioned $T$- and $\mu$-independence of $\Delta n_{\sigma}(x,t)$ for massless Dirac fermions. Indeed for massive fermions with a non-linear dispersion relation, Eq.(\ref{eq-axial-anomaly})  would display an additional (non-anomalous) term, proportional to the mass and dependent on the electronic state\cite{bertlmann}.
\subsection{Photoexcited local energy correlations}
Let us now consider the electronic correlations. While in Ref.\cite{dolcini_2016} we have focussed mostly on the photoexcited momentum distribution, which is a correlation of electron operators at different space points and at the same time $t^\prime=t$, here we shall focus on the local correlation (i.e. at equal space point $x^\prime=x$) at different times, described by  
\begin{eqnarray}
\mathcal{G}_\sigma(t_2,t_1;x)   \doteq  e^{-\frac{i{\rm e}}{\hbar}  \int_{t_2}^{t_1}   V(x^\prime,t)  \, d t^\prime} \,
  \left\langle {\psi}_{\sigma}^\dagger(x,t_1)   \psi_{\sigma}(x,t_2)   \right\rangle     \hspace{0.5cm}  \sigma=\uparrow,\downarrow  \,,\label{es-mat-def} 
\end{eqnarray}
where  the phase pre-factor involving the scalar potential $V$   is nothing but the Wilson line (\ref{WL-def}) in the particular case of equal space point correlations $x^\prime=x$, and guarantees the  gauge invariance of the Eq.(\ref{es-mat-def}).  
By Fourier transforming Eq.(\ref{es-mat-def}) with respect to the time difference $t^\prime=t_1-t_2$ and   the average time $t=(t_1+t_2)/2$, one obtains the  correlation  in the energy domain,  
\begin{eqnarray}  
 \mathcal{G}_\sigma(E,\tilde{E};x)     =\frac{v_F}{2\pi \hbar}\int\!\!\!\int dt \,  dt^\prime e^{-i \frac{E t^\prime}{\hbar}} e^{-i \frac{\tilde{E} t}{\hbar}} \mathcal{G}_{\sigma}(t-\frac{t^\prime}{2},t+\frac{t^\prime}{2};x) \,\,. \label{nu_{pm}-def-omega-space}
\end{eqnarray}
Inserting   the exact space-time evolution obtained in Eq.(\ref{Psi-sol-gen}) for the electron field operator into Eq.(\ref{es-mat-def}) and (\ref{nu_{pm}-def-omega-space}), one expresses the energy correlation as
\begin{equation}
\mathcal{G}_{\uparrow,\downarrow}(E,\tilde{E};x) = \frac{1}{4\pi \hbar}  \int_{-\infty}^{+\infty} \!\!\! \!\!dt \, e^{-i \tilde{E} t/\hbar}    \left(-i \lim_{a \rightarrow 0} \int_{-\infty}^{+\infty}
 \!\! \frac{e^{\mp i  \Delta\phi^{\rm es}_{\uparrow,\downarrow}(t,t^\prime;x)} e^{-i(E -\mu) t^\prime/\hbar  }  }{l_T \sinh[\pi (v_F t^\prime- i a)/l_T]}\, d( v_F t^\prime) \right) , \label{Gomega-tildeomega}
\end{equation} 
where  $l_T=\hbar v_F/k_BT$ is the thermal lengthscale related to the equilibrium temperature $T$, whereas $a=\hbar v_F/E_c$ is a short distance related to the ultra-violet energy cutoff $E_c$. Furthermore, the dimensionless quantity
\begin{eqnarray}
\Delta \phi^{\rm es}_{\uparrow,\downarrow}(t,t^\prime;x) \doteq   \phi_{\uparrow,\downarrow}(x,t+\frac{t^\prime}{2}) -\phi_{\uparrow,\downarrow}(x,t-\frac{t^\prime}{2}) \pm \frac{{\rm e}}{\hbar}\int_{t-\frac{t^\prime}{2}}^{t+\frac{t^\prime}{2}} \!\!\!V(x,  t^\prime) \, dt^\prime  \label{Deltaphig-es-def} 
\end{eqnarray}
denotes the  {\it gauge invariant} phase difference at equal space (es) points $x=x^\prime$,  where  the second term on the right-hand side compensates for the gauge dependence of the first term $\phi_{\uparrow,\downarrow}$, given by Eqs.(\ref{phi-up})-(\ref{phi-dn}).

In particular, for  the initial equilibrium state the correlation (\ref{es-mat-def}) is given by 
\begin{eqnarray}
\mathcal{G}^\circ_\sigma(t_2,t_1;x) =\langle {\psi^\circ_\sigma}^\dagger (x,t_1) {\psi^\circ_\sigma}^{}(x,t_2)  \rangle_\circ = -i \,  \frac{e^{i \mu(t_1-t_2)/\hbar  }}{2 l_T \sinh\left[\pi   v_F(t_1-t_2-\frac{i\hbar}{E_c}) /l_T \right]} \,,\hspace{1cm} \label{corr-0}
\end{eqnarray}
and depends only on the time difference. Thus,  once inserted  into Eq.(\ref{nu_{pm}-def-omega-space}), it straightforwardly yields the diagonal equilibrium energy correlation (\ref{Gcirc}), which vanishes for any $\tilde{E} \neq 0$.

In contrast,  when the time-dependent  electric pulse is applied, the out of equilibrium correlation (\ref{es-mat-def}) does not necessarily depend on the  difference $t^\prime=t_1-t_2$ between the two time arguments, due to the phase difference (\ref{Deltaphig-es-def}) and the Wilson prefactor. Thus, the photoexcited energy distribution, i.e. the deviation of Eq.(\ref{Gomega-tildeomega}) from the equilibrium value (\ref{Gcirc}),  also contain `off-diagonal' terms $\tilde{E} \neq 0$,   and can be expressed as   
\begin{eqnarray}  
\lefteqn{\Delta \mathcal{G}_{\uparrow,\downarrow}(E,\tilde{E};x)  = \mp  \frac{1}{4\pi \hbar} \frac{1}{l_T }   \int_{-\infty}^{+\infty} \!\!\! dt\, e^{-i \tilde{E} t/\hbar} \times } & &   \nonumber \\
& & \times  \int_{-\infty}^{+\infty} \, \, \frac{\sin\left[\Delta\phi^{\rm es}_{\uparrow,\downarrow}(t,t^\prime;x) \pm (E -\mu) t^\prime/\hbar \right]\, \mp \sin\left[ (E -\mu) t^\prime/\hbar\right]}{\sinh[\pi v_F t^\prime/l_T]}\, d(v_F t^\prime) \quad. \label{DeltaG_{pm}-def} 
\end{eqnarray}
The photoexcited energy distribution is obtained as the diagonal term $\tilde{E} \rightarrow 0$ of the photoexcited energy correlations, i.e. 
\begin{equation}
\label{Deltanu-def}
\Delta \nu_{\uparrow,\downarrow}(E;x) \doteq \Delta\mathcal{G}_{\uparrow,\downarrow}(E,\tilde{E}=0;x)   \,\,\, \,\quad.
\end{equation}

\section{Explicit results for specific cases}
\label{sec:4}
\subsection{Plane wave pulse}
In order to appreciate the role of the finite duration of the applied electric pulse, let us first consider the case where a plane wave is applied for a time duration $\tau$
\begin{equation} 
E(x,t)= E_0\,  \theta(\frac{\tau}{2}-t) \theta(\frac{\tau}{2}+t) \, \cos(\frac{\Omega x}{c}) \, \cos(\Omega t)  \quad, \label{E-planewave-finitetau}
\end{equation}
where $\Omega$ and $\tau$ are the frequency  the duration of the pulse, respectively, and $\theta$ denotes the Heaviside function. 
Applying the general result Eqs.(\ref{Deltan_{up}_res-gen})-(\ref{Deltan_{dn}_res-gen}) to Eq.(\ref{E-planewave-finitetau}), and focussing on the times $t>\tau/2$ after the end of the pulse, one finds 
\begin{equation}
\Delta n_{\uparrow,\downarrow}(x,t) =\displaystyle \pm \frac{{\rm e}E_0}{2\pi \hbar \Omega} \, \left(\sum_{r=\pm} \frac{ \sin\left[  \Omega  \tau (1+r \frac{v_F}{c})/2\right]}{ (1+r\frac{v_F}{c})} \right)    \cos \left[  \frac{\Omega}{c} (x \mp v_F  t) \right]   \hspace{0.5cm} t>\frac{\tau}{2} \,\, .\label{Deltan_{pm}-planewave-finitetau}
\end{equation}
We can now consider two limits of Eq.(\ref{Deltan_{pm}-planewave-finitetau}).
 For a long pulse and/or high frequency ($\Omega \tau \gg 1)$   one straightforwardly sees that the photoexcited density profiles vanishes,
\begin{equation}
\Delta n_{\uparrow,\downarrow}(x,t)  = \displaystyle \pm \frac{{\rm e}E_0}{2\hbar\Omega} \, \cos \left[  \frac{\Omega}{c} (x\mp v_F  t) \right]    \,  \left(\delta(1+\frac{v_F}{c})+ \delta(1-\frac{v_F}{c}) \right)  \,\, =0 \quad. \label{Deltan_{pm}-planewave-finitetau-hih}
\end{equation}
In this case the electron system probes the whole plane wave both in space and time, and momentum and energy  conservation 
laws lead to a vanishing response: there cannot be intra-branch transitions because of the difference between the electron Fermi velocity $v_F$ and the speed of light $c$. 

In contrast, in the case of a short pulse and/or low frequency ($\Omega \tau \ll 1)$ one finds
\begin{equation}
\displaystyle  
\Delta n_{\uparrow,\downarrow}(x,t) = \displaystyle \pm \frac{{\rm e}E_0 \tau}{2\pi \hbar} \,   \cos \left[  \frac{\Omega}{c} (x \mp v_F  t) \right]    \label{Deltan_{pm}-planewave-finitetau-low}
\end{equation}
In this situation only the momentum   $q=\Omega/c$ is conserved, whereas energy is not, so that the electron density propagates oscillating in time with a  frequency 
\begin{equation}
\Omega_{el}=\Omega v_F/c
\end{equation}
lower than the frequency $\Omega$ of the electromagnetic wave.
In particular, denoting by~$L$  the size of the system, in the customary situation $\Omega L/c \ll 1$,  one obtains
\begin{equation}
\Delta n_{\uparrow,\downarrow}(x,t)\simeq  \displaystyle \pm \frac{{\rm e}E_0 \tau}{2\pi \hbar} \,   \cos \left[  \frac{v_F}{c} \Omega  t \right]     \label{Deltan_{pm}-planewave-finitetau-low-2}
\end{equation}
i.e. the electron system sees a spatially uniform field that oscillates in time with a lower frequency $\Omega_{el}=\Omega v_F/c$, and the charge distribution is also uniform. 
\subsection{Gaussian electric pulse}
Let us now consider  the case of an electric field that is localized both in time and space. In particular, a Gaussian electric pulse is described by
\begin{equation}\label{Egauss-cos}
E(x,t)=E_0 \, e^{-\frac{x^2}{2 \Delta^2}} \, e^{-\frac{t^2}{2 \tau^2}} \quad,
\end{equation}
where $\Delta$ and $\tau$ denote its space extension and   time duration around the space and time origin, respectively, and $E_0$ is its amplitude. Here below we present the result for the case (\ref{Egauss-cos}), focussing on the density space profile  and the energy distribution of the photoexcited wave packets. 
Substituting the pulse (\ref{Egauss-cos}) into Eqs.(\ref{Deltan_{up}_res-gen})-(\ref{Deltan_{dn}_res-gen}), one obtains
\begin{eqnarray}
\displaystyle  \Delta n_{\uparrow,\downarrow}(x,t)=  \pm \frac{{\rm e} E_0}{2 \pi \hbar}\frac{D}{v_F} \sqrt{\frac{\pi}{2}} e^{-\frac{(x \mp v_F t)^2}{2(\Delta^2+(v_F \tau)^2)} }    \left[ 1+{\rm Erf}\left(\frac{D}{\sqrt{2}}\left(\pm \frac{x}{\Delta^2}+\frac{t}{v_F \tau^2} \right)\!\right)\right]     \label{Deltan_{pm}-Egauss-pulse}
\nonumber
\end{eqnarray}
where
\begin{equation}\label{D-def}
D  \doteq    \frac{v_F \tau \Delta}{\sqrt{\Delta^2+(v_F \tau)^2}}
\end{equation}
is an effective length scale involving both the space extension $\Delta$ and the time duration $\tau$ of the pulse.  At long times ($t\gg \tau$) and away from the region of the applied pulse ($\sigma x \gg \Delta$), the argument of the Error function is large, and one can find the asymptotic expression    
\begin{equation}
\Delta n_{\uparrow,\downarrow}(x,t)  \simeq   \displaystyle  \pm \frac{{\rm e} E_0}{2 \pi \hbar}\frac{\tau \Delta\sqrt{2 \pi}}{\sqrt{\Delta^2+(v_F \tau)^2}}  e^{-\frac{(x \mp v_F t)^2}{2(\Delta^2+(v_F \tau)^2)} } \, , \label{Deltan_{pm}-Egauss-pulse-zero-omega-long-times}
\end{equation}
which describes two spin-polarised photoexcited wave packets propagating rightwards and leftwards, respectively. Notice that  the shape of the electron densities $\Delta n_{\uparrow,\downarrow}$ is Gaussian. However, the space extention 
\begin{equation}
\Delta_{el}=\sqrt{\Delta^2+(v_F \tau)^2}
\label{Delta-el}
\end{equation}
depends on {\it both}   the space extension $\Delta$ and the time duration $\tau$ of the electric pulse, and is bigger than $\Delta$. 
\subsection{Spatial $\delta$-pulses}
In the particular limit where the electric pulse is applied over a short region, $\Delta\ll v_F \tau$, Eq.(\ref{Egauss-cos}) can be treated as a spatial $\delta$-pulse, $E(x,t)= \mathcal{E}_0 \, \delta(x)\exp(- t^2/2 \tau^2)$   upon identifying $\mathcal{E}_0=E_0  \Delta  \sqrt{2\pi}$. In this case the spatial extension (\ref{Delta-el}) of the photoexcited electron density profile is only given by $\Delta_{el} \simeq v_F \tau$, where  $\tau$ is the pulse duration. This is a particular case of a spatial $\delta$-pulse. Let us thus analyze this situation in more general terms,  allowing for a generic time dependence $\mathcal{V}(t)$ for the spatial $\delta$-pulse centered   around $x=0$,
\begin{equation}\label{E-delta-space}
E(x,t)=  \delta(x)  \,\mathcal{V}(t)  \hspace{2cm}
\end{equation}
where  $[\mathcal{V}]={\rm voltage}$. Among   all possible gauges   reproducing the electric pulse Eq.(\ref{E-delta-space}), two are worth being mentioned,
\begin{eqnarray}
& \mbox{(pure $A$-gauge)}&\hspace{2cm}  \left\{ \begin{array}{ccl}
V&=&0 \\
A(x,t)&=&-c \, \delta(x)  \int_{-\infty}^t  \!\mathcal{V}(t^\prime)\, dt^\prime  
\end{array}\right.\label{A-delta-space}
\\ \nonumber \\
& \mbox{(pure $V$-gauge)}&\hspace{2cm}  \left\{ \begin{array}{ccl}
V(x,t)&=& \theta(-x)\, \mathcal{V}(t)  \\
A&=&0 \quad \quad.\\
\end{array}\right.\label{V-delta-space}
\end{eqnarray}

The photoexcited density profiles are obtained by inserting Eq.(\ref{E-delta-space}) into the general result (\ref{Deltan_{up}_res-gen})-(\ref{Deltan_{dn}_res-gen}), obtaining
\begin{equation}
\Delta n_{\uparrow,\downarrow}(x,t) =  \pm  \frac{{\rm e} }{2\pi \hbar v_F} \theta(\pm x) \mathcal{V}\left(t \mp \frac{x}{v_F}\right)  \quad,
 \label{Deltan_{pm}-delta-space}
\end{equation}
whence we see that the {\it spatial shape} of the two counter-propagating spin polarized wavepackets is determined by the {\it time profile} $\mathcal{V}$ of the pulse.
 
The phases induced by the electromagnetic field are obtained from  the general results (\ref{phi-up})-(\ref{phi-dn}) and, in particular for the two gauges (\ref{V-delta-space}) and (\ref{A-delta-space}), one gets
\begin{eqnarray}
& \mbox{(pure $A$-gauge)}&\hspace{0.4cm} 
\phi_{\uparrow,\downarrow}(x,t) = \displaystyle -\frac{{\rm e} }{\hbar} \theta(\pm x) \int^{t \mp \frac{x}{v_F}}_{-\infty}  \mathcal{V}(t^{\prime\prime})\, dt^{\prime\prime}  \hspace{0.5cm}  \label{phi_{pm}-delta-space-A}  \\
& \mbox{(pure $V$-gauge)}&\hspace{0.4cm}   \phi_{\uparrow,\downarrow}(x,t) =  - \frac{{\rm e}}{\hbar} \theta(\pm x) \int_{-\infty}^{t\mp \frac{x}{v_F}} \!\!\!   \mathcal{V}(t^\prime)    \, dt^\prime     \,\mp \frac{{\rm e}}{\hbar}  \theta(-x)   \int_{-\infty}^{t}  \mathcal{V}(t^\prime)    \, dt^\prime \hspace{0.5cm}\label{phi_{pm}-delta-space-V} 
\end{eqnarray}
while the gauge invariant  phase difference Eq.(\ref{Deltaphig-es-def})  determining the photoinduced energy correlations through Eq.(\ref{DeltaG_{pm}-def}) is given by
\begin{equation}
 \Delta \phi^{\rm es}_{\uparrow,\downarrow}(t,t^\prime;x)  =\displaystyle   -\frac{{\rm e}}{\hbar} \theta(\pm x) \int_{t\mp \frac{x}{v_F}-\frac{t^\prime}{2}}^{t\mp \frac{x}{v_F}+\frac{t^\prime}{2}} \,    \mathcal{V}(t^{\prime\prime})    \, dt^{\prime\prime}  \quad.   \label{phi_{pm}-es-delta-space} 
\end{equation}

\subsubsection{The case of a localized Lorentzian pulse}
Let us consider as a specific example the case of spatial $\delta$-pulse (\ref{E-delta-space}) with a Lorentzian time shape,  
\begin{equation}\label{Lor}
\mathcal{V}(t)=\frac{2\hbar u_0}{{\rm e}} \frac{\tau}{t^2+\tau^2}
\end{equation}
where $u_0$ is a dimensionless amplitude parameter and $\tau$ is the  duration timescale. 

\begin{figure}
\resizebox{0.7\columnwidth}{!}{\includegraphics{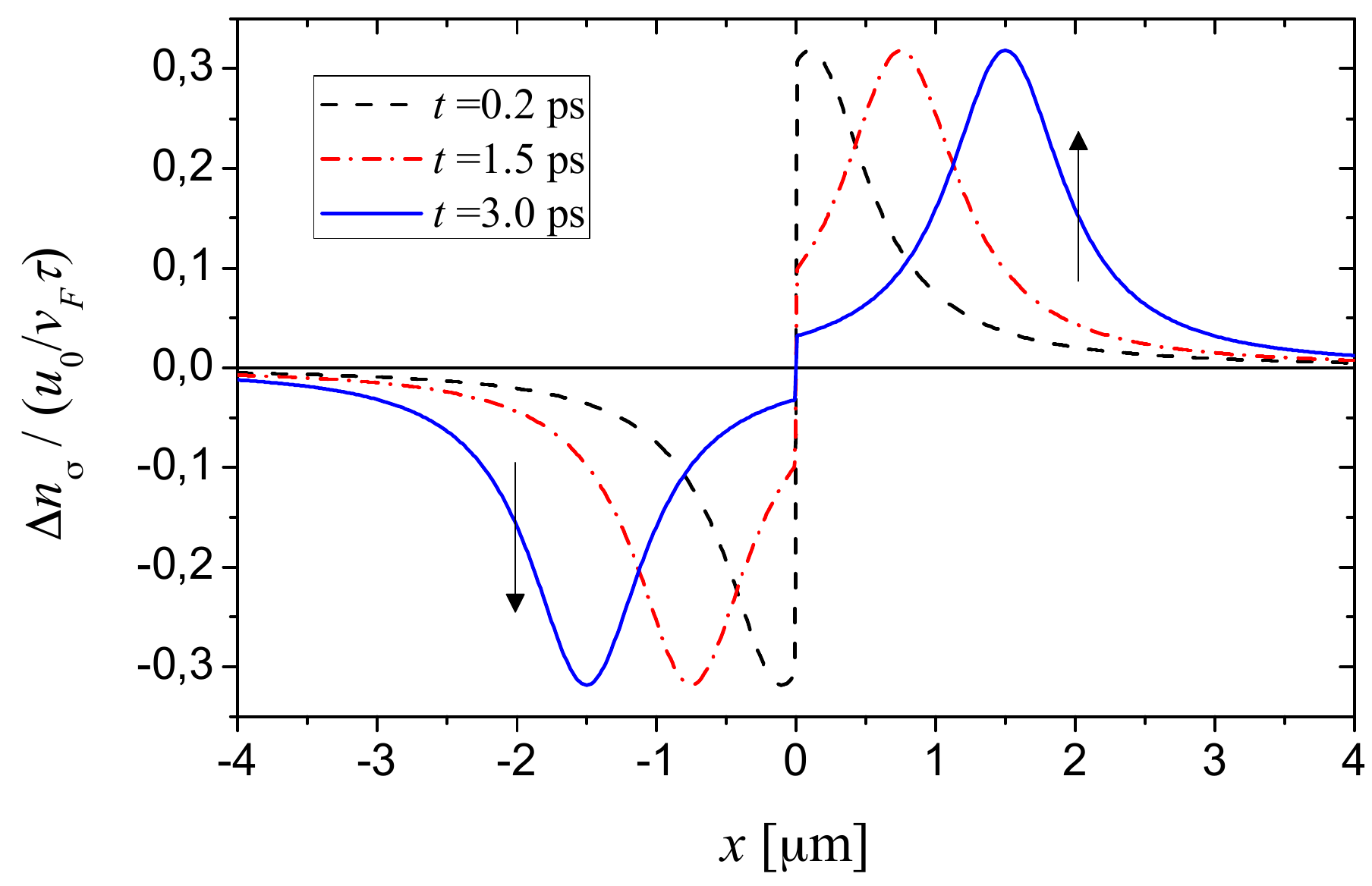} }
\caption{Photoexcited electron density profiles, generated by a spatial $\delta$-pulse with Lorentzian time profile characterized by $\tau=1{\rm ps}$ [see Eqs.(\ref{E-delta-space}) and (\ref{Lor})], and  are plotted as a function of the distance along the QSH edge from the location of the pulse $x=0$, at different time snapshots, $t=0.2\,{\rm ps}$ (black dashed curve), $t=1.5\,{\rm ps}$ (red dash-dotted curve)  and $t=3\,{\rm ps}$ (blue solid curve). The two wavepackets propagate in opposite directions with opposite spin. Their space profile maintains unaltered without dispersion and is independent of the temperature of the initial equilibrium state, due to the linearity of the  QSH helical edge states spectrum. The value $v_F=5 \times 10^5\,{\rm m/s}$ is taken for the QSH edge Fermi velocity.}
\label{fig:1}       
\end{figure}

The photoexcited density profiles are straightforwardly obtained from Eq.(\ref{Deltan_{pm}-delta-space}) 
\begin{equation}
\Delta n_{\uparrow,\downarrow}(x,t) =  \pm  \theta(\pm x)  \frac{u_0}{\pi  v_F \tau} \frac{(v_F \tau)^2}{\left(x\mp v_F \tau\right)^2+(v_F\tau)^2} \label{Deltan_{pm}-delta-space-lor}
\end{equation}
and are depicted in Fig.\ref{fig:1}, where the different curves refer to various time snapshots. 
We emphasize that, from the general result of Ref.\cite{dolcini_2016} presented in Sec.\ref{sec:3:1}, the density profile (\ref{Deltan_{pm}-delta-space-lor}) is temperature independent for arbitrary $u_0$. This was also confirmed for the case   $u_0=1$ in Ref.\cite{moskalets2017}.\\

As one can see, two spin-polarised wavepackets are gradually generated at the origin $x=0$, where the pulse is applied, and start to travel in opposite directions  with a Lorentzian spatial shape  with a spatial extension $v_F \tau$.  We emphasize that i) the  photoexcited density profile preserves its shape without any dispersion and ii)  is independent of the temperature and the chemical potential of the initial equilibrium state\cite{dolcini_2016}. Both these features are due to the linearity of the massless Dirac spectrum of QSH edge states. Indeed similar features have been recently found in metallic carbon nanotubes\cite{rosati2015} also described by a similar model.\\

\begin{figure}
\center
\resizebox{0.7\columnwidth}{!}{\includegraphics{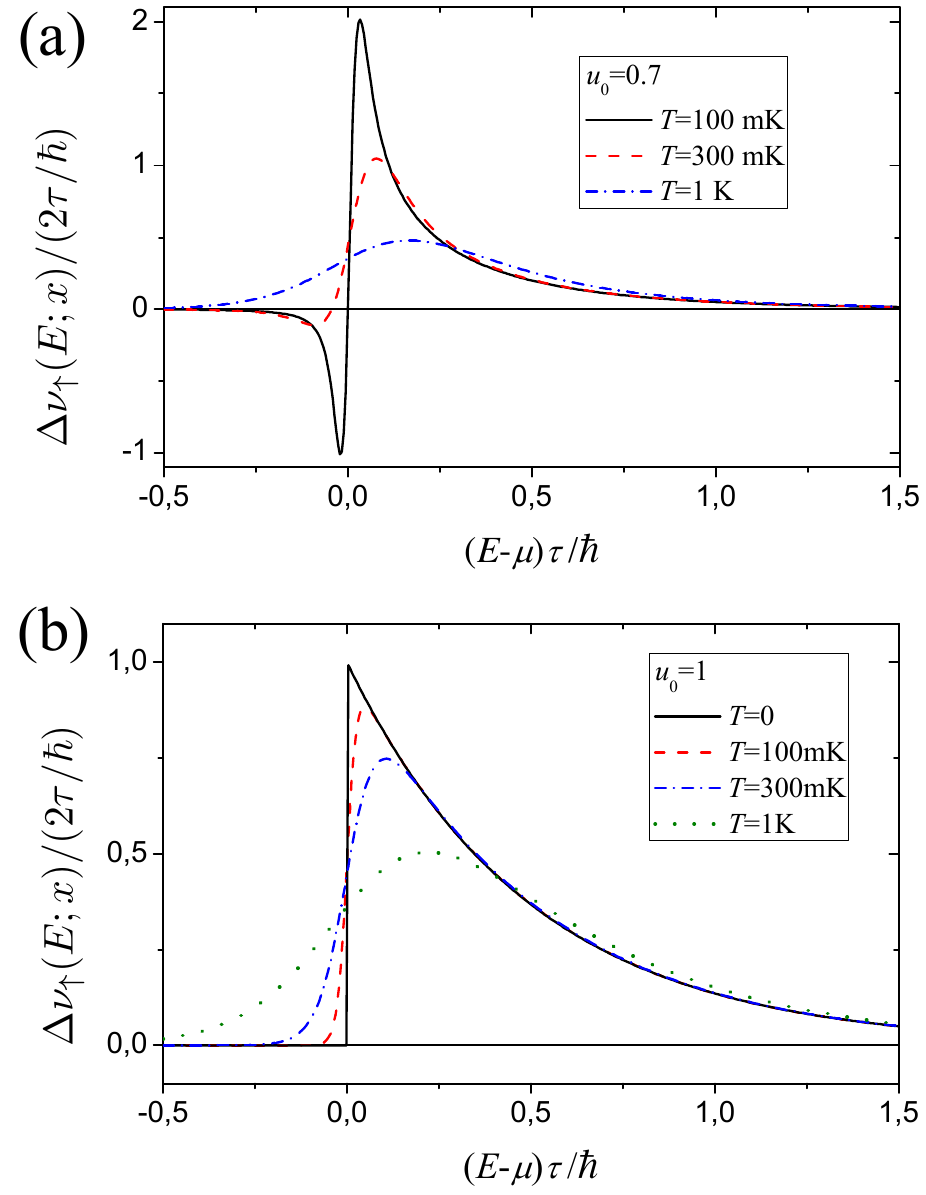} }
\caption{The energy distribution  Eq.(\ref{Deltanu-def}) of the spin-$\uparrow$ electrons photoexcited by a space $\delta$-pulse is plotted as a function of the energy deviation from the initial equilibrium chemical potential $\mu$, in units of $2\tau/\hbar$. The   time shape of the pulse is Lorentzian, Eq.(\ref{Lor}), with time parameter  $\tau=1\,{\rm ps}$ and amplitude parameter $u_0=0.7$ [panel (a)], and $u_0=1$  [panel (b)].  The various curves refer to different temperatures, whose values are indicated in the legenda, and are essentially independent of the  location $x$, due to the fact that the pulse is applied over a very narrow $\delta$-like region.   In both cases it appears that $\Delta\nu_{\uparrow}(E;x)$   depends on temperature, quite differently from the spatial density profile (see Fig.\ref{fig:1}) and that   the integral of $\Delta\nu_{\uparrow}(E;x)$ over energy is not vanishing: the photoexcitation process does not simply redistribute  electrons from below to above the Fermi level, a clear signature of the chiral anomaly effect characterizing QSH edge states. In particular, the case (b) describes purely particle excitations, also known as Levitons, whose zero temperature expression is given by Eq.(\ref{Deltanu-Lev}). }
\label{fig:2}       
\end{figure}

Let us now consider  the  energy correlations induced by the photoexcitation. To this purpose, we first derive from Eq.(\ref{phi_{pm}-es-delta-space}) the  gauge invariant phase difference, obtaining
\begin{eqnarray}
\Delta \phi^{\rm es}_{\uparrow,\downarrow}(t,t^\prime;x)=\displaystyle   -2 u_0 \theta(\pm x) \left( \arctan\left(\frac{t+ \frac{t^\prime}{2}}{\tau}\right)-     \, \arctan\left(\frac{t- \frac{t^\prime}{2}}{\tau}\right)\right) \label{Deltaphi-es-delta-space-lor}
\end{eqnarray}
and then insert it in Eq.(\ref{DeltaG_{pm}-def}). From Eq.(\ref{Deltaphi-es-delta-space-lor}) we note that
$
\Delta \phi^{\rm es}_{\uparrow}(t,t^\prime;x)=\Delta \phi^{\rm es}_{\downarrow}(t,t^\prime;-x)
$
and  we can then deduce from Eq.(\ref{DeltaG_{pm}-def})  the general relation
\begin{equation}\label{DeltaG+DeltaG-}
\Delta\mathcal{G}_{\downarrow}(E-\mu, \tilde{E};x) =-\Delta\mathcal{G}_{\uparrow}(\mu-E, \tilde{E};-x) \quad.
\end{equation}
It is therefore enough to compute  only  $\Delta\mathcal{G}_{\uparrow}$. In particular, we shall focus on the diagonal limit $\tilde{E}\rightarrow 0$ [see Eq.(\ref{Deltanu-def})], which describes the photoexcited energy distribution, i.e. the deviation from the equilibrium Fermi distribution (\ref{Gcirc}).  The result is plotted in Fig.\ref{fig:2}  as a function of energy deviation from the equilibrium Fermi level,  for two values of the  Lorentzian amplitude parameter, $u_0=0.7$  in panel (a) and $u_0=1$  in panel (b). Two important aspects emerge. 
The first one can be deduced by inspecting the various curves, which describe the behavior of $\Delta\nu_\uparrow(E;x)$ for different values of temperatures: The photoexcited energy distribution $\Delta \nu_\uparrow(E;x)$ does depend  on the temperature $T$ and on the chemical potential $\mu$ of the initial equilibrium state. On the one hand, this can easily be understood from the fact that the equilibrium  distribution characterizing the initial state occupancy determines which states are   available for  the photoexcitation process. On the other hand, this is quite different from the photoexcited density profiles $\Delta n_{\uparrow,\downarrow}(x,t)$ [see Eqs.(\ref{Deltan_{up}_res-gen})-(\ref{Deltan_{dn}_res-gen})], which are temperature independent. This means that, for QSH edge states, despite the redistribution induced by the photoexcitation process on the energy depends on the initial state, it always occurs in such a way that the corresponding density profile is insensitive to such initial state. For the sake of clarity, it is worth recalling that the photoexcited energy distribution is not  the  Fourier transform of the photoexcited density space profile (which a local quantity in space and time), as the former takes into account also the correlations at {\it different} times at that space point. This is why the essential difference in terms of the temperature behavior emerges.

The second noteworthy aspect emerging from Fig.\ref{fig:2}(a) is that, although the curves in general display a depletion (hole excitations) below the Fermi energy and and enhancement (particle excitations) above the Fermi level, the integral of $\Delta\nu_\uparrow(E;x)$ over energy is typically not zero. This means that, in general, the photoexcitation process in QSH edge states does not simply generate particle-hole excitations by reshuffling states from below to above the Fermi level and viceversa. Indeed, besides this particle-hole contribution, there exists a net charge `creation' effect for spin-$\uparrow$ electrons. This appears clearly in the case $u_0=1$, shown in Fig.\ref{fig:2}(b), where the photoexcitation  generates    purely particle excitations above the Fermi level, without holes.  Such effective creation effect is a result of the infinite number of states characterizing the ground state of Dirac fermions:  the applied electric pulse   pulls up states from the depth of the Fermi sea so that, in comparison with the equilibrium state, a net charge is generated. This `creation'
 is of course compensated by the annihilation effect of spin-$\downarrow$ electrons, $\Delta\nu^\circ_{\downarrow}(E-\mu;x) =-\Delta\nu^\circ_{\uparrow}(\mu-E;-x)$, as can be deduced by Eq.(\ref{DeltaG+DeltaG-}), consistently with the fact that no net charge can be created by an electromagnetic pulse (the careful treatment of gauge invariance precisely guarantees that the total charge of the system is conserved). This is the gist of the chiral anomaly effect characterizing massless Dirac fermions\cite{nielsen1983}: Although the two spin components are not coupled directly by the electromagnetic field, they are coupled indirectly via the infinite depth of the Fermi sea. This physically means that, above the QSH bulk gap,  the edge channels get coupled through the bulk states of the QSH quantum well. A similar  situation occurs in   one-dimensional ballistic channels of a 2DEG, where only near the Fermi level can  the parabolic spectrum be well approximated by two decoupled linear branches of right- and left-moving electrons. However, near the band bottom such decoupling is not well defined and the hole left at the bottom of the Fermi sea by a (say) photoexcited right-moving electron  can be occupied by a left-moving electron, which in turn leaves a hole at the other Fermi point\cite{levitov2006}. \\

The peculiarity of the case $u_0=1$ shown in  Fig.\ref{fig:2}(b)  was first noticed by Levitov and coworkers\cite{levitov1996,levitov1997,levitov2006}, and can be rigorously proven by noticing that, in such case, the exponential factor $e^{-i\Delta \phi_\uparrow(t,t^\prime;x)}$ appearing in Eq.(\ref{Gomega-tildeomega})  reduces to
\begin{equation}
\left. e^{-i\Delta \phi_\uparrow(t,t^\prime;x)}\right|_{u_0=1}=\prod_{r=\pm} \frac{t^\prime+2 r t-2 i\tau}{t^\prime+2 r t+2i\tau}\quad,
\end{equation}
which is an analytical function in the upper half-plane of $t^\prime$. Thus,  when $E<\mu$, a closing of the   integral contour of Eq.(\ref{Gomega-tildeomega}) in such half-plane only leaves the equilibrium contribution. A non vanishing  photoexcited local energy distribution  can only be found  above the equilibrium Fermi level (purely particle excitation). Explicitly, at zero temperature and $u_0=1$ one finds  
\begin{equation}
\begin{array}{l}
\left. \Delta \nu_{\uparrow}(E;x) \right|_{T=0\,;\, u_0=1}=   \theta(x)  \, \theta(E-\mu) \,    \frac{2\tau}{\hbar} \,  e^{-2 |E-\mu|\tau/\hbar}   \,  
\end{array}\label{Deltanu-Lev}
\end{equation}
and, more in general, for the photoexcited energy correlations
\begin{equation}
\begin{array}{ll}
\left. \Delta \mathcal{G}_{\uparrow,\downarrow}(E,\tilde{E};x) \right|_{T=0\,;\, u_0=1}=  & \pm\theta(\pm x)  \, \theta(\pm(E-\mu)) \,  e^{\mp i  \tilde{E}x/\hbar v_F}  \frac{2\tau}{\hbar} \,\cdot  e^{-2 |E-\mu|\tau/\hbar}  \times \nonumber \\
&   \times  \, \left(\mbox{sgn}[2|E -\mu|+  \tilde{E}]+\mbox{sgn}[2|E -\mu|-  \tilde{E}]\right)/2
\end{array}\label{DeltaG-Lev}
\end{equation}
Note that Eq.(\ref{DeltaG-Lev}) is independent of the position $x$.
The existence of such minimal excitations, also called Levitons, is currently on the spotlight in electron quantum optics\cite{moskalets2014,flindt2015,moskalets2016,moskalets2017,martin-sassetti_PRL_2017,martin-sassetti_PRB_2017}, especially after their experimental observation  in ballistic channels of a 2DEG\cite{glattli2013,glattli_PRB_2013,glattli2014,glattli2017} and the proposal of their detection in QH systems\cite{glattli_PRB_2013b}.

%

\section{Interplay between photoexcitation and Rashba spin-orbit coupling}
\label{sec:5}
In QSH systems, besides the bulk spin-orbit coupling underlying the topological transition and giving rise to the very existence of the edge states,  Rashba spin-orbit coupling (RSOC) can emerge along the edge    either because of disorder effects, due to the random ion distribution in the heterostructure doping layers and to the random bonds at the  quantum well interfaces~\cite{sherman_PRB_2003,sherman_APL_2003}, or  by intentional deformation of  the boundary curvature\cite{entin_2001,ojanen_2011,ortix_2015,gentile_2015,cuoco-2016} or also  by the electric field itself applied to local metallic gate electrodes, e.g. to generate the electric pulse for the photoexcitation~\cite{molenkamp_2006,niu_2013,park_2013,wolosyn_2014}.  For these reasons, we wish to address the interplay between RSOC and photoexcitation.
In the presence of RSOC, the electronic term appearing  in (\ref{Hfull}) becomes
\begin{equation}\label{Hcirc-2}
\hat{\mathcal{H}}_\circ=\hat{\mathcal{H}}_{kin}+\hat{\mathcal{H}}_R
\end{equation} 
where the Rashba-coupling term
\begin{eqnarray}
\label{HR}
\hat{\mathcal{H}}_R = \frac{1}{\hbar} \int   dx \, \Psi^\dagger(x) \,   \frac{1}{2} \left\{ \alpha_R(x) \, , {p}_x   \right\}  \sigma_2\, \Psi(x) 
\end{eqnarray}
is characterised by a profile $\alpha_R(x)$ and depends linearly on the momentum $p_x$. As a consequence, the current operator $\hat{J}$ appearing in the electromagnetic coupling (\ref{Hem}) is modified into
\begin{eqnarray}
\hat{J}&=&v_F \Psi^\dagger(x)   \left(\sigma_3+\frac{\alpha_R(x)}{\hbar v_F }   \sigma_2  \right)\Psi(x)=  v_F\left[ \hat{n}_\uparrow-\hat{n}_\downarrow +\frac{i  \alpha_R(x)}{\hbar v_F}(\psi^\dagger_\downarrow\psi^{}_\uparrow-\psi^{\dagger}_\uparrow\psi^{}_\downarrow) \right] \label{J-def-bis}
\end{eqnarray}
in order to ensure charge conservation. 
The Hamiltonian (\ref{Hfull}) can be compactly rewritten as
$
\hat{\mathcal{H}}  =  \int dx \, \Psi^\dagger(x) \, H(x) \, \Psi(x) \label{H}
$, 
where the  first-quantized Hamiltonian density is
\begin{eqnarray} \label{Hmat} 
H(x)   
&=&  \frac{v_F}{2} \left\{ \sigma_3 +\tan\theta_R(x) \sigma_2 \, ,  {p}_x - \frac{{\rm e}}{c} A(x,t) \right\} \, \, \, +{\rm e} V(x,t) \sigma_0\,\,,
\end{eqnarray}
with $\sigma_0$ denoting the $2 \times 2$ identity matrix and $\theta_R \in [-\pi/2; + \pi/2]$ is the Rashba angle, defined as
\begin{equation} \label{thetaR-def}
\theta_R(x) \doteq \arctan \frac{\alpha_R(x)}{\hbar v_F}  \quad.
\end{equation}
Due to the RSOC, the Hamiltonian (\ref{Hmat}) is no longer diagonal, and   the  dynamics of the spin components $\psi_\uparrow$ and $\psi_\downarrow$ is   coupled. This suggests that the  field $\Psi$, i.e. the basis of spin components, is not the most suitable one to obtain the dynamical evolution. Instead, it is worth  switching to another basis, by re-expressing the electron field spinor $\Psi$ as a rotation around $\sigma_1$ by the space-dependent Rashba angle~$\theta_R(x)$  
\begin{equation}\label{Psi-from-X-compact}
\Psi(x)    \doteq    e^{+\frac{i}{2} \sigma_1\theta_R(x)} \, {\rm X}(x)   
\end{equation}
where
\begin{equation}\label{X-def}
{\rm X}(x)=\left(\begin{array}{l}\chi_{+}(x) \\ \\ \chi_{-}(x) \end{array} \right)
\end{equation}
is called the ``chiral'' field spinor.  The rotation (\ref{Psi-from-X-compact}) 
enables one to rewrite the Hamiltonian~(\ref{Hfull}) as
$
\hat{\mathcal{H}} =\int dx \,\Psi^\dagger(x) H(x) \Psi^{}(x)     =  \int dx \,{\rm X}^\dagger(x) H_\chi(x) {\rm X}(x) 
$,
where  
\begin{eqnarray}
 H_\chi(x)  
=  \frac{1}{2}\left\{ v(x)    ,\, {p}_x  -\frac{{\rm e}}{c}  A(x,t) \right\} \sigma_3 \, +{\rm e}   V(x,t) \sigma_0 \hspace{0.5cm} 
\label{Hchi}
\end{eqnarray}
is the   Hamiltonian  of massless Dirac fermions travelling with a {\it space-dependent} velocity  profile
\begin{equation}\label{v(x)-def}
v(x)=\frac{v_F}{\cos\theta_R(x)} = v_F \sqrt{1+\left(\frac{\alpha_R(x)}{\hbar v_F}\right)^2} \, \, \ge v_F 
\end{equation}
and exposed to the electromagnetic field. 
In the chiral basis (\ref{X-def}) the RSOC, encoded in the   profile (\ref{v(x)-def}), always {\it increases} the velocity with respect to the bare value of Fermi velocity $v_F$, regardless of the sign of $\alpha_R$.  
Furthermore, in the chiral basis the density (\ref{n-def}) and the current density (\ref{J-def}) acquire  simple expressions, namely
$
\hat{n}(x)=  \,{\rm X}^\dagger(x)   \,   {\rm X}(x) =\hat{n}_{+}+\hat{n}_{-}$ and $\hat{J}(x) =  v(x) \,{\rm X}^\dagger(x) \sigma_3 \,   {\rm X}(x) =v(x) (\hat{n}_{+}-\hat{n}_{-} )$,
with $\hat{n}_\pm \doteq \chi^\dagger_\pm \chi^{}_\pm$.
Finally we emphasise that, since $H_\chi$ is  diagonal  in the chiral basis [see Eq.(\ref{Hchi})], the two components~$\chi_\pm$ of  Eq.(\ref{X-def}) are dynamically {\it decoupled}, even when the electromagnetic field is applied and the RSOC is present.  
This feature implies that,  in striking contrast with the original spin components $\psi_\uparrow$ and $\psi_\downarrow$, which have a well defined propagation direction only away from the Rashba interaction region, the chiral components $\chi_{+}$ and $\chi_{-}$ describe genuine right-moving and left-moving electrons, respectively, even in the regions where Rashba interaction is present. This is the origin of the term ``chiral'' and the reason for considering ${\rm X}$ as the ``natural basis'' for Rashba-coupled states.  

In the chiral basis it is now easy to generalize the results obtained in Sec.\ref{sec:2} to the case of an inhomogeneous velocity. The dynamical evolution of the electron field operator is given by
\begin{eqnarray}
 \left(\partial_t \pm v(x)\partial_x \pm\frac{1}{2} \partial_x v(x)   \right) \chi_\pm =    - \frac{i{\rm e}}{\hbar} \left(    V(x,t)    \mp  \frac{v(x)}{c}\,A(x,t) \, \right)  \chi_\pm \quad, \label{eq-chi-with-field} 
\end{eqnarray}
whose  solution   is\cite{dolcini_2017}
\begin{equation}\label{chipm-sol}
\chi^{}_\pm(x,t) =   e^{\pm i \phi_\pm(x,t)}\, \chi^\circ_\pm(x,t) \quad.
\end{equation}
Here 
\begin{equation}\label{chipm0-sol}
\chi^\circ_\pm(x,t)=  \frac{1}{\sqrt{2 \pi \hbar \,v(x)}} \int dE \, e^{ - i \frac{E}{\hbar}    \left(t \mp \int_{x_{r}}^x \frac{d x^{\prime \prime}}{v(x^{\prime \prime})} \right)}\,\hat{c}_{E\pm}  
\end{equation}
  is a solution   for the pulse-free case, Eq.(\ref{Hcirc-2}), where the exponential phases clearly show that $\chi^\circ_\pm$ are genuine right- and left-moving electrons, respectively, propagating with the inhomogeneous velocity (\ref{v(x)-def}), while $\hat{c}^{}_{E+}$ and $\hat{c}^{}_{E-}$ denote fermionic operators for right- and left-moving electrons at the energy $E$, fulfilling $\{ \hat{c}^{}_{E\pm} \, , \hat{c}^\dagger_{E^\prime \pm} \}=\delta(E-E^\prime)$, and $x_{r}$ denotes an arbitrarily fixed reference point, such as the space origin or the geometrical center of the RSOC profile.
Furthermore, in Eq.(\ref{chipm-sol})   
\begin{equation}\label{phipm}
\begin{array}{l}
\phi_+(x,t) = \displaystyle {\frac{{\rm e}}{\hbar c} \int_{-\infty}^x \!\!\!\! dx^\prime  \left( A-\frac{c}{v(x^\prime)}V\right) (x^\prime,t-\int_{x^\prime}^{x} \frac{dx^{\prime \prime}}{v(x^{\prime\prime})}) }   \\  \\
\phi_-(x,t) = \displaystyle  {\frac{{\rm e}}{\hbar c} \int_{x}^{\infty}  \!\!\!\! dx^\prime   \left( A+\frac{c}{v(x^\prime)}V\right) (x^\prime,t-\int_{x}^{x^\prime} \frac{dx^{\prime \prime}}{v(x^{\prime\prime})}) }   
\end{array}  
\end{equation}
describe the phases induced by the electromagnetic field, and generalize the expressions (\ref{phi-up})-(\ref{phi-dn}) obtained without RSOC.  The result (\ref{chipm-sol}) thus   describes {\it exactly} --within the assumption of independent electrons-- the electron dynamics in the presence of both the RSOC and the electromagnetic field.  \\

\begin{figure}
\center
\resizebox{0.7\columnwidth}{!}{\includegraphics{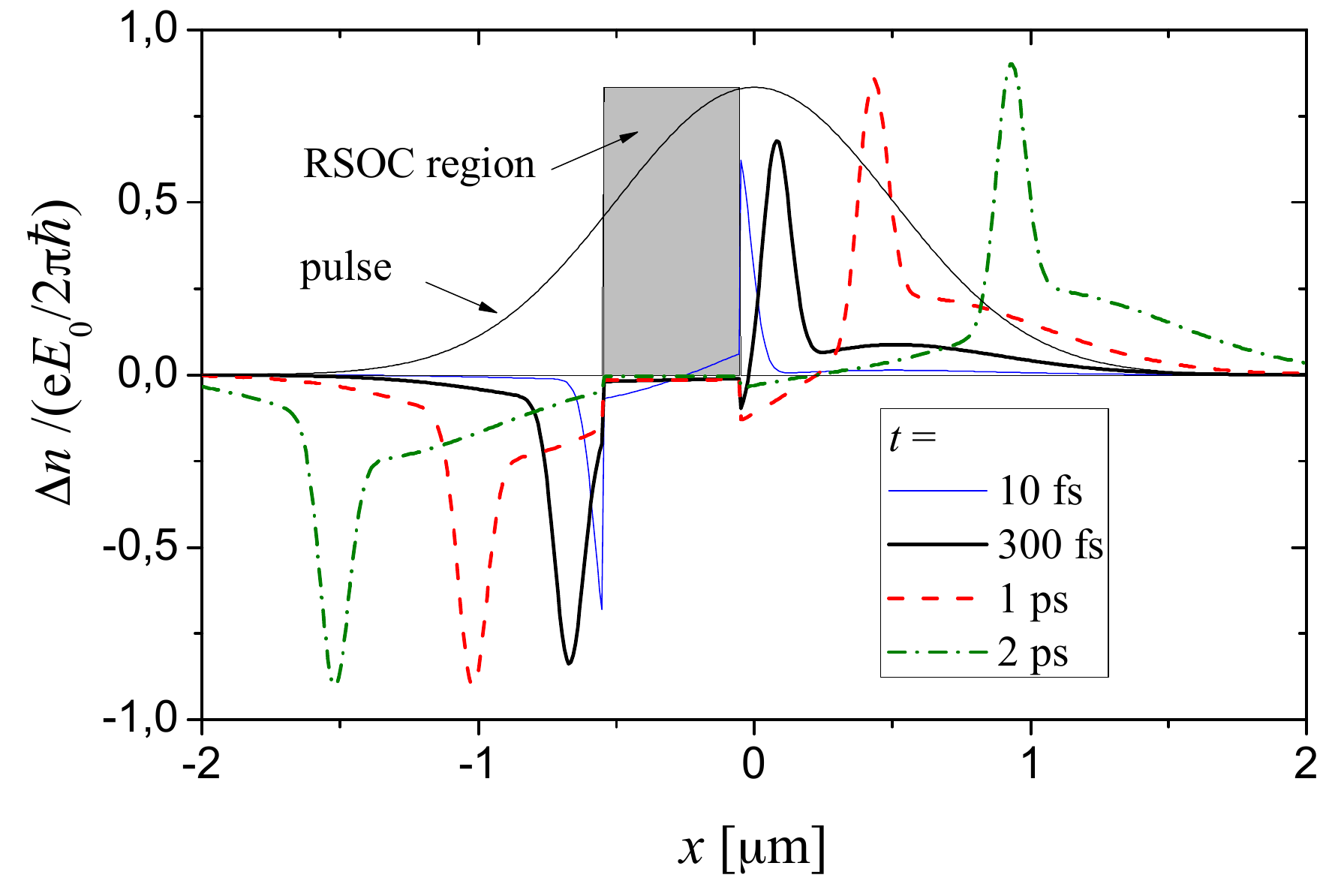} }
\caption{The photoexcitation process induced by a Gaussian electric pulse (\ref{Egauss-cos}) (thin black solid line) partially overlapping with  a RSOC region (grey area): the profile of the photoexcited densities is plotted as a function of the  spatial coordinate $x$ along the QSH edge, at different time snapshots.  The parameters of the Gaussian pulse (\ref{Egauss-cos}) are $\Delta=500\,{\rm nm}$ and $\tau=0.1\,{\rm ps}$. The inhomogeneous Fermi velocity (\ref{v(x)-def}) induced by the RSOC leads to a fuzzy shape of the wavepackets,  resulting from a combination of the photoexcitation inside and outside the RSOC region, where electrons have different   Fermi velocities.  The RSOC can thus be exploited to tailor the shape of photoexcited wavepackets.}
\label{fig:3}       
\end{figure}

Proceeding in a similar way as was done in Sec.\ref{sec:3}, one obtains the expressions for the right- and left-moving photoexcited density profiles $\Delta  {n}_\pm$, namely
\begin{eqnarray}
\Delta  {n}_+(x,t) &=& \displaystyle  +\frac{{\rm e}}{2\pi\hbar v(x)} \int_{-\infty}^x dx^\prime E(x^\prime,t-\int_{x^\prime}^{x} \frac{dx^{\prime \prime}}{v(x^{\prime \prime})})    \label{Delta-n_+}\\
\Delta  {n}_-(x,t)
&=& \displaystyle-\frac{{\rm e}}{2\pi\hbar v(x)} \int_{x}^{\infty}  dx^\prime E(x^\prime,t-\int_{x}^{x^\prime} \frac{dx^{\prime \prime}}{v(x^{\prime \prime})})   \quad, \label{Delta-n_-}
\end{eqnarray}
which are shown in Fig.\ref{fig:3}, where the interplay of a Gaussian  electric pulse (\ref{Egauss-cos}) with a  RSOC region (grey area) is described. Because of the inhomogeneous Fermi velocity (\ref{v(x)-def}) induced by the RSOC, the electrons photoexcited inside and outside the RSOC region have different  Fermi velocities,   generating the fuzzy shape of the wavepackets. This interplay shows that the RSOC can   be exploited to tailor the shape of photoexcited wavepackets.
Other examples and a thorough discussion of these effects can be found in Ref.\cite{dolcini_2017}.
 
\section{Proposal for experimental realisations}
\label{sec:6}
 Two main  realisations of QSH systems presently exist, namely   in HgTe/CdTe
\cite{bernevig_science_2006,konig_2006,molenkamp-zhang_jpsj,roth_2009,brune_2012} and in InAs/GaSb~\cite{liu-zhang_2008,knez_2007,knez_2014,spanton_2014} quantum wells. In their topological phase, conducting helical edge states  appear and exhibit a linear dispersion with a Fermi velocity $v_F \simeq 5 \times 10^5 {\rm m/s}$ and $v_F \simeq 2 \times 10^4 {\rm m/s}$,  respectively~\cite{molenkamp-zhang_jpsj,knez_2014},   within a bulk gap   $E_g \sim 30 \,{\rm meV}$. The phase breaking length $L_\phi$, i.e. the length scale for which the  analysis carried out in this paper is valid, is of the order of a few micrometers at Kelvin temperatures. 

A localised electromagnetic pulse can be generated with two techniques. 
The first one is the use of near field scanning optical microscopy operating in the illumination mode: an optical fiber with a thin aperture of tens of nanometers, positioned near the edge, excites a strong electric field at the tip apex~\cite{novotny_review,koch1997,novotny2003,nomura2011,nomura2015}. With this sophisticated technique one obtains  localised pulses, whose spatial center can also be easily  displaced.  The second approach to create a localised electromagnetic pulse is somewhat  more straightforward: It consists in utilising side finger gate electrodes, deposited close to a boundary of the QSH bar  and biased by   time-dependent voltages experienced by the electrons in the edge\cite{dolcini_2012}, similarly to what has been proposed for a 2DEG~\cite{bocquillon2014,levitov2006,glattli2013,glattli_PRB_2013,glattli2014,glattli2017}. In this case the spatial extension of the electric pulse is determined by the lateral width of the finger electrode,  $\sim 100 {\rm nm}$, which can be biased by a time-dependent gate voltage $\mathcal{V}(t)$. Note that  the pure photoexcitation process does not involve any electron tunneling from the finger electrodes, differently from the case of electron pumps\cite{feve2007,ritchie_2007,feve2011,kataoka2013,kataoka2015,janssen2016,sassetti2017}.  
The recent advances in pump-probe experiments and photo-current spectroscopy~\cite{bocquillon2014,glattli2010,glattli2013,lesueur2009,holleitner2014,fujisawa2014,holleitner2015,jarillo-herrero2016,fujisawa2017}, make the time-resolved  detection of the photoexcited wave packets realistically accessible nowadays.

\section{Conclusions}
\label{sec:7}
In conclusion, we have shown that, while in the customary far field regime vertical electric dipole transitions are forbidden in QSH edge states,   an electric pulse  localised in space and/or time and applied at the edge of a QSH system  can photoexcite electron wavepackets by intra-branch electrical transitions, without invoking the bulk states or the Zeeman coupling. Several interesting features are found. First, the   wavepackets are  spin-polarised and  propagate in opposite directions. Their density profile depends linearly on the amplitude of the electric pulse,  is independent of the    initial equilibrium temperature and  does not  disperse during propagation [see Fig.\ref{fig:1}]. This result,  quite different from the one obtained in usual Schr\"odinger systems described by a parabolic spectrum,   is exact and does not rely on any linear response approximation. It  is due to the linearity of the spectrum.
Secondly, we have analyzed the energy correlations and in particular  the  photoinduced energy distribution. We have shown that such quantity does depend on temperature and  on the amplitude of the applied pulse in a highly non-linear way [see Fig.\ref{fig:2}]. Furthermore  in energy domain the photoexcitation process does not merely amount to a redistribution of states from below to above the Fermi level, rather it also involves a net creation of charge on one branch (compensated by the annihiliation on the other branch). This effect, which is the gist of the chiral anomaly   characterizing massless Dirac electrons, is particularly evident in the case of a Lorentzian pulse with amplitude parameter $u_0=1$ [see Fig.\ref{fig:2}(b)] that describes minimal particle excitations known as Levitons\cite{levitov1996,levitov1997,levitov2006}. These results may pave the way to the observation of Levitons in QSH, whose existence has been so far experimentally proven in ballistic channels in 2DEG
\cite{glattli2013,glattli_PRB_2013,glattli2014,glattli2017}. 
Also, we have discussed the effects of Rashba spin-orbit coupling on the photoexcitation process, showing that the former can be exploited to tailor the shape of photoexcited wavepackets [see Fig.\ref{fig:3}]. Finally, we have proposed possible experimental realizations. These results support the idea that QSH edge states might be successfully exploited in the near future as a promising alternative  platform for   electron quantum optics\cite{bercioux_2013,buttiker_2013,martin_2014,recher_2015,sassetti_2016,dolcini_2016,dolcini_2017}, which is nowadays mostly limited to quantum Hall systems~\cite{feve2007,roulleau2008,glattli2010,feve2011,bocquillon2013,bocquillon_science_2013,kataoka2013,bocquillon2014,martin_2014b,feve2015,kataoka2015,janssen2016} with the unavoidable drawback of the strong magnetic fields. In contrast, time-reversal topological insulators, which are based on spin-orbit coupling, are immune to such drawback and offer the additional possibility of generating spin-polarised electron wave packets.    \\

\section*{Acknowledgment}
Inspiring discussions with E. Bocquilllon, R. C. Iotti, and A. Montorsi are greatly acknowledged. The authors contributed equally to  this manuscript.

%
 



\begin{thebibliography}{}

\bibitem{bertoni2000} A. Bertoni, P. Bordone, R. Brunetti, C. Jacoboni, S. Reggiani, Phys. Rev. Lett.  \textbf{84},  (2000) 5912
\bibitem{ritchie_2007} M. D. Blumenthal, B. Kaestner, L. Li, S. Giblin, T. J. B. M. Janssen, M. Pepper,
D. Anderson, G. Jones, and  D. A. Ritchie, Nature Phys.  \textbf{3},  (2007) 343
\bibitem{bocquillon2012} E. Bocquillon, F. D. Parmentier, C. Grenier, J.-M. Berroir,  P. Degiovanni, B. Pla\c{c}ais, A. Cavanna, Y. Jin, and G. F\`eve, Phys. Rev. Lett. \textbf{108}, (2012) 196803








\bibitem{feve2007} G. F\`eve,  A. Mah\'e, J.-M. Berroir,  T. Kontos,  B. Pla\c{c}ais,  D. C. Glattli, A. Cavanna,  B. Etienne,  Y. Jin, Science \textbf{316},  (2007) 1169

\bibitem{roulleau2008} P. Roulleau, F. Portier, P. Roche, A. Cavanna, G. Faini, U. Gennser, and D. Mailly, 
Phys. Rev. Lett. \textbf{100}, (2008) 126802 

\bibitem{glattli2010}  A. Mah\'e, F. D. Parmentier, E. Bocquillon, J.-M. Berroir, D. C. Glattli, T. Kontos, B. Pla\c{c}ais, G. F\`eve, A. Cavanna, and Y. Jin, Phys. Rev. B \textbf{82}, (2010) 201309(R) 


\bibitem{feve2011} Ch. Grenier, R. Herv\'e, E. Bocquillon, F. D. Parmentier, B. Pla\c{c}ais, J. M. Berroir, G. F\`eve and P. Degiovanni, New J. Phys. \textbf{13}, (2011) 093007 

\bibitem{bocquillon2013} E. Bocquillon, V. Freulon, J.-M. Berroir, P. Degiovanni, B. Pla\c{c}ais, A. Cavanna, Y. Jin, and G. F\`eve, Nature Commun. \textbf{4},  (2013) 1839 
\bibitem{bocquillon_science_2013} E. Bocquillon,  V. Freulon,  J.-M Berroir,  P. Degiovanni,  B. Pla\c{c}ais,  A. Cavanna,  Y. Jin,  G. F\`eve, Science {\bf 339}, (2013) 1054 
\bibitem{kataoka2013} J. D. Fletcher,  P. See,  H. Howe,  M. Pepper,  S. P. Giblin,  J. P. Griffiths,  G. A. C. Jones, I. Farrer,  D. A. Ritchie,  T. J. B. M. Janssen, and M. Kataoka, Phys. Rev. Lett. \textbf{111}, (2013) 216807 

\bibitem{bocquillon2014} E. Bocquillon, V. Freulon, F. D. Parmentier, J.-M Berroir, B. Pla\c{c}ais, C. Wahl, J. Rech, T. Jonckheere, T. Martin, C. Grenier, D. Ferraro, P. Degiovanni, G. F\`eve, Ann. Phys. (Berlin) \textbf{526},  (2014) 1




\bibitem{martin_2014b}   C. Wahl, J. Rech, T. Jonckheere, and T. Martin, Phys. Rev. Lett. \textbf{112},  (2014) 046802 

\bibitem{feve2015} V. Freulon, A. Marguerite, J.-M. Berroir, B. Pla\c{c}ais, A. Cavanna, Y. Jin, and G. F\`eve,   Nature Commun. \textbf{6},  (2015) 6854

\bibitem{kataoka2015} J. Waldie,  P. See,  V. Kashcheyevs,  J. P. Griffiths,  I. Farrer,  G. A. C. Jones,  D. A. Ritchie, T. J. B. M. Janssen,  and M. Kataoka, Phys. Rev. B \textbf{92},  (2015) 125305


\bibitem{janssen2016} M. Kataoka,  N. Johnson,  C. Emary, P. See,  J. P. Griffiths,  G. A. C. Jones,  I. Farrer, D. A. Ritchie,  M. Pepper, and T. J. B. M. Janssen, Phys. Rev. Lett. \textbf{116}, (2016) 126803 


\bibitem{kane-mele2005a} C. L. Kane, E. J. Mele, Phys. Rev. Lett. \textbf{95}, (2005) 146802 
\bibitem{kane-mele2005b} C. L. Kane, E. J. Mele, Phys. Rev. Lett. \textbf{95},  (2005) 226801
\bibitem{bernevig_science_2006} B. A. Bernevig, T. L. Hughes, and S.-C. Zhang, Science \textbf{314},  (2006) 1757

\bibitem{konig_2006}  M. K\"onig, S. Wiedmann, C. Br\"une, A. Roth, H. Buhmann, L. W. Molenkamp, X.-L.  Qi, and S.-C. Zhang, Science \textbf{318}, (2006)  766
\bibitem{molenkamp-zhang_jpsj} M. K\"onig, H. Buhmann, L. W. Molenkamp, T. L. Hughes, C.-X. Liu, X.-L. Qi, and S.-C. Zhang, J. Phys. Soc. Jpn. \textbf{77}, (2008) 031007 
\bibitem{roth_2009} A. Roth, C.~Br\"une, H. Buhmann, L. W. Molenkamp, J.~ Maciejko, X.-L. Qi, and S.-C. Zhang, Science \textbf{325}, (2009)  294 
\bibitem{brune_2012} C. Br\"une,	A. Roth, H. Buhmann,	E. M. Hankiewicz, L. W. Molenkamp, J. Maciejko, X.-L. Qi, and S.-C. Zhang, Nature Phys. \textbf{8},  (2012) 485
 
 
\bibitem{liu-zhang_2008} C. Liu, T. L. Hughes,  X.-L. Qi, K. Wang, and S.-C. Zhang, Phys. Rev. Lett. \textbf{100},  (2008) 236601
\bibitem{knez_2007} I. Knez, R.-R. Du, and G. Sullivan, Phys. Rev. Lett. \textbf{107},  (2011) 136603
\bibitem{knez_2014} I. Knez, C. T. Rettner, S.-H. Yang, and S. S. P. Parkin, L. Du,  R.-R. Du, and 
G. Sullivan, Phys. Rev. Lett. \textbf{112},  (2014) 026602
\bibitem{spanton_2014} E. M. Spanton, K C. Nowack, L. Du, G. Sullivan,  R.-R. Du, and K. A. Moler, Phys. Rev. Lett. \textbf{113},  (2014) 026804


\bibitem{bercioux_2013} A. Inhofer  and D. Bercioux, Phys. Rev. B \textbf{88},  (2013) 235412
\bibitem{buttiker_2013} P. P. Hofer, M. B\"uttiker, Phys. Rev. B \textbf{88},  (2013) 241308
\bibitem{martin_2014}  D. Ferraro, C. Wahl, J. Rech, T. Jonckheere, and T. Martin, Phys. Rev. B  \textbf{89},  (2014) 075407 
\bibitem{recher_2015} A. Str\"om, H. Johannesson, and P. Recher, Phys. Rev. B \textbf{91},  (2015) 245406
\bibitem{sassetti_2016} A. Calzona, M. Acciai, M. Carrega, F. Cavaliere, and M. Sassetti, Phys. Rev. B \textbf{94}, (2016)  035404
\bibitem{dolcini_2016} F. Dolcini, R. C. Iotti, A. Montorsi, and F. Rossi, Phys. Rev. B \textbf{94}, (2016) 165412 
\bibitem{dolcini_2017} F. Dolcini, Phys. Rev. B \textbf{95}, (2017) 085434



\bibitem{cayssol2012} B. D\'ora, J. Cayssol, F. Simon, and R. Moessner, Phys. Rev. Lett. \textbf{108},  (2012) 056602
\bibitem{artemenko2013} S. N. Artemenko, and V. O. Kaladzhyan,  JETP Lett. \textbf{97}, (2013)  82 
\bibitem{dolcetto-sassetti2014} G. Dolcetto, F. Cavaliere, and M. Sassetti, Phys. Rev. B \textbf{ 89},  (2014) 125419

\bibitem{artemenko2015} V. Kaladzhyan, P. P. Aseev,  and S. N. Artemenko, Phys. Rev. B \textbf{92}, (2015) 155424 







\bibitem{levitov1996} L. S. Levitov, H. Lee, and G. B. Lesovik, J. Math. Phys. \textbf{37},  (1996) 4845
\bibitem{levitov1997} D. A. Ivanov, H.W. Lee, and L. S. Levitov, Phys. Rev. B \textbf{56},  (1997) 6839
\bibitem{levitov2006} J. Keeling, I. Klich, and L. S. Levitov, Phys. Rev. Lett. \textbf{97},  (2006) 116403
\bibitem{moskalets2014} M. Moskalets, Phys. Rev. B \textbf{89} (2014), 045402 
\bibitem{flindt2015} D. Dasenbrook, and C. Flindt, Phys. Rev. B \textbf{92}, (2015) 161412 
\bibitem{moskalets2016} M. Moskalets, Phys. Rev. Lett. \textbf{117}, (2016) 046801 
\bibitem{moskalets2017} M. Moskalets, Low Temp. Phys. \textbf{43} (2017), 865 
\bibitem{martin-sassetti_PRL_2017} J. Rech, D. Ferraro, T. Jonckheere, L. Vannucci, M. Sassetti, and T. Martin,  Phys. Rev. Lett. \textbf{118} (2017), 076801
\bibitem{martin-sassetti_PRB_2017} L. Vannucci, F. Ronetti, J. Rech, D. Ferraro, T. Jonckheere, T. Martin, and M. Sassetti,  Phys. Rev. B \textbf{95} (2017), 245415 



\bibitem{glattli2013} J. Dubois, T. Jullien, F. Portier, P. Roche, A. Cavanna, Y. Jin, W. Wegscheider, P. Roulleau, and D. C. Glattli, Nature \textbf{502},  (2013) 659
\bibitem{glattli_PRB_2013} J. Dubois, T. Jullien, C. Grenier, P. Degiovanni, P. Roulleau, and D. C. Glattli, Phys. Rev. B \textbf{88},  (2013) 085301
\bibitem{glattli2014}   T. Jullien, P. Roulleau, B. Roche, A. Cavanna, Y. Jin, and  D. C. Glattli, Nature \textbf{514}, (2014) 603 
\bibitem{glattli2017}  D. C. Glattli, P. S. Roulleau, Phys. Status Sol. \textbf{254} (2017), 1600650 






\bibitem{adler1969} S. L. Adler, Phys. Rev. \textbf{177}, (1969)  2426 
\bibitem{bell-jackiw1969} J.S. Bell and R. Jackiw, Nuovo Cim. \textbf{A 60}, (1969) 47 
\bibitem{nielsen1983} H. B. Nielsen, N. Ninomiya, Phys. Lett. \textbf{B130},  (1983) 389
\bibitem{bertlmann} R. A. Bertlmann, {\it Anomalies in quantum field theory}, (Clarendon Press, Oxford, 1996)  

\bibitem{burkov2012} A. A. Zyuzin and A. A. Burkov, Phys. Rev. B \textbf{86},  (2012) 115133
\bibitem{li2013} H.-J. Kim,  K.-S. Kim, J.-F. Wang,  M. Sasaki,  N. Satoh, A. Ohnishi,  M. Kitaura,  M. Yang,  and L. Li, Phys. Rev. Lett \textbf{111},  (2013) 246603
\bibitem{wishvanath2014} S. A. Parameswaran, T. Grover,  D. A. Abanin, D. A. Pesin,  and A. Vishwanath, Phys. Rev. X \textbf{4},  (2014) 031035
\bibitem{takane2016} Y. Takane, J. Phys. Soc. Jpn \textbf{85},  (2016) 013706
 

 


\bibitem{chen2014} Z. K. Liu, B. Zhou, Y. Zhang, Z. J. Wang, H. M. Weng,
D. Prabhakaran, S.-K. Mo, Z. X. Shen, Z. Fang, X. Dai, Z. Hussain, and Y. L. Chen, Science \textbf{343}, (2014) 864 
\bibitem{hasan2015a} S.-Y. Xu, I. Belopolski, N. Alidoust, M. Neupane, G.Bian, C. Zhang, R. Sankar, G. Chang, Z. Yuan, C.-C. Lee, S.-M. Huang, H. Zheng, J. Ma, D. S. Sanchez, B.Wang, A. Bansil, F. Chou, P. P. Shibayev, H. Lin, S. Jia, and M. Z. Hasan, Science \textbf{349}, (2015)  613 
\bibitem{hasan2015b} S.-Y. Xu, I. Belopolski, D. S. Sanchez, C. Zhang, G.
Chang, C. Guo, G. Bian, Z. Yuan, H. Lu, T.-R. Chang, P.
P. Shibayev, M. L. Prokopovych, N. Alidoust, H. Zheng,
C.-C. Lee, S.-M. Huang, R. Sankar, F. Chou, C.-H. Hsu,
H.-T. Jeng, A. Bansil, T. Neupert, V. N. Strocov, H. Lin,
S. Jia, and M. Z. Hasan, Science \textbf{349},  (2015) 622
\bibitem{ong2015} J. Xiong, S. K. Kushwaha, T. Liang, J. W. Krizan, M.
Hirschberger, W. Wang, R. J. Cava, and N. P. Ong, Science
\textbf{350},  (2015) 413

\bibitem{trauzettel2016} C. Fleckenstein, N. Traverso Ziani,  and B. Trauzettel,  Phys. Rev. B  \textbf{94}, (2016)  241406(R) 





\bibitem{rosati2015} R. Rosati, F. Dolcini and F. Rossi, Appl. Phys. Lett. \textbf{106},  (2015) 243101


\bibitem{glattli_PRB_2013b}   Ch. Grenier, J. Dubois, T. Jullien, P. Roulleau,  D. C. Glattli, and P. Degiovanni, Phys. Rev. B \textbf{88} (2013), 085302 
 


\bibitem{sherman_PRB_2003} E. Ya. Sherman, Phys. Rev. B {\bf 67},  (2003) 161303
\bibitem{sherman_APL_2003} E. Ya. Sherman, Appl. Phys. Lett. {\bf 82}, (2003) 209 


\bibitem{entin_2001} M. V. Entin and L. I. Magarill, Phys. Rev. B {\bf 64},  (2001) 085330
\bibitem{ojanen_2011} J. I. V\"ayrynen and T. Ojanen, Phys. Rev. Lett. {\bf  106},  (2011)   076803
\bibitem{ortix_2015} C. Ortix, Phys. Rev. B {\bf 91},  (2015) 245412
\bibitem{gentile_2015} P. Gentile, M. Cuoco,  and C. Ortix, Phys. Rev. Lett. {\bf 115}, (2015) 256801 
\bibitem{cuoco-2016} Z.-J. Ying, P. Gentile, C. Ortix, and M. Cuoco, Phys. Rev. B {\bf 94},  (2016) 081406(R).


\bibitem{molenkamp_2006} J. Hinz, H. Buhmann, M. Sch\"afer, V. Hock, C. R. Becker, and L. W. Molenkamp, Semi. Sci. Tech. {\bf 21},  (2006) 501
\bibitem{niu_2013} Z. Qiao, X. Li, W. K. Tse, H. Jiang,  Y. Yao,  and Q. Niu, Phys. Rev. B {\bf 87},  (2013) 125405
\bibitem{park_2013} Y. H. Park, S.-H. Shin, J. D. Song, J. Chang, S. H. Han, H.-J. Choi, H. C. Koo, Solid-State Electron. {\bf 82},  (2013) 23
\bibitem{wolosyn_2014} P. W\'{o}jcik, J. Adamowski, B. J. Spisak, and M. Wo{\l}oszyn, J. Appl. Phys. {\bf 115},  (2014) 104310



\bibitem{novotny_review} L. Novotny, and S. J. Stranick, Ann. Rev. Phys. Chem. \textbf{57},  (2006) 303
\bibitem{koch1997} B. Hanewinkel, A. Knorr, P. Thomas, and S.W. Koch, Phys. Rev. B \textbf{55}, (1997) 13715 
\bibitem{novotny2003} A. Hartschuh, E. J. S\'anchez,  X. S. Xie,  and L. Novotny, Phys. Rev. Lett. \textbf{90}, (2003)  095503
\bibitem{nomura2011} H. Ito,  K. Furuya,  Y. Shibata,  S. Kashiwaya,  M. Yamaguchi,  T. Akazaki,  H. Tamura,  Y. Ootuka,  and S. Nomura, Phys. Rev. Lett. \textbf{107},  (2011) 256803
\bibitem{nomura2015} S. Mamyouda,  H. Ito, Y. Shibata, S. Kashiwaya, M. Yamaguchi, T. Akazaki, H. Tamura, Y. Ootuka, and S. Nomura, Nanolett. \textbf{15},  (2015) 4127

\bibitem{dolcini_2012} F. Dolcini Phys. Rev. B \textbf{85}, (2012)  033306 

\bibitem{sassetti2017}   M. Acciai, A. Calzona, G. Dolcetto, T. L. Schmidt,  and M. Sassetti, Phys. Rev. B \textbf{96}, 075144 (2017) 

\bibitem{lesueur2009} C. Altimiras, H. le Sueur, U. Gennser, A. Cavanna, D. Mailly and F. Pierre, Nature Phys. \textbf{6},  (2009) 34





\bibitem{holleitner2014} A. Brenneis, L. Gaudreau, M. Seifert, H. Karl, M. S. Brandt, H. Huebl,
J. A. Garrido, F. H. L. Koppens, and A. W. Holleitner, Nature Nanotech. \textbf{10},  (2014) 135
\bibitem{fujisawa2014}  H. Kamata, N. Kumada, M. Hashisaka,    K. Muraki, and T. Fujisawa, Nature Nanotech. \textbf{9}, (2014) 177 
\bibitem{holleitner2015} C. Kastl, C. Karnetzky, H. Karl, and A. W. Holleitner, Nature Commun. \textbf{6}, (2015) 6617 
\bibitem{jarillo-herrero2016} A. Woessner, P. Alonso-Gonz\'alez, M. B. Lundeberg, Y. Gao, J. E. Barrios-Vargas,
G. Navickaite, Q. Ma, D. Janner, K. Watanabe, A. W. Cummings, T. Taniguchi,
V. Pruneri, S. Roche, P. Jarillo-Herrero, James Hone, Rainer Hillenbrand, and F. H. L. Koppens, Nature Commun. \textbf{7},  (2016) 10783
\bibitem{fujisawa2017}  M. Hashisaka,  N. Hiyama, T. Akiho, K. Muraki, and T. Fujisawa, Nature Phys. \textbf{13}, (2017) 559 







 
%
%



\end{thebibliography}
\end{document}